\documentclass[iop]{emulateapj}
\RequirePackage{epstopdf}
\usepackage{multirow}
\RequirePackage{graphicx}

\newcommand{\water}{H$_2$O}

\newcommand{\teff}{T$_{\mathrm{eff}} $ }

\newcommand{\um}{$\mu$m}

\slugcomment{AJ, in press}

\shorttitle{Prospecting in late-type dwarfs}
\shortauthors{Mann et al.}

\begin{document}

\title{Prospecting in late-type dwarfs: \\ a calibration of infrared and visible spectroscopic metallicities of late-K and M dwarfs spanning $1.5$ dex}

\author{Andrew W. Mann\altaffilmark{1}, John M. Brewer\altaffilmark{2}, Eric Gaidos\altaffilmark{3}, S\'{e}bastien L\'{e}pine\altaffilmark{4}, and Eric J. Hilton\altaffilmark{1,3} }
  
\altaffiltext{1}{Institute for Astronomy, University of Hawai'i, 2680 Woodlawn Dr, Honolulu, HI 96822} 
\altaffiltext{2}{Department of Astronomy, Yale University, New Haven, CT 06511, USA} 
\altaffiltext{3}{Department of Geology \& Geophysics, University of Hawai'i, 1680 East-West Road, Honolulu, HI 96822} 
\altaffiltext{4}{Department of Astrophysics, American Museum of Natural History, New York, NY 10024}

\begin{abstract}
Knowledge of late K and M dwarf metallicities can be used to guide planet searches and constrain planet formation models. However, the determination of metallicities of late-type stars is difficult because visible wavelength spectra of their cool atmospheres contain many overlapping absorption lines, preventing the measurement of equivalent widths. We present new methods, and improved calibrations of existing methods, to determine metallicities of late-K and M dwarfs from moderate resolution ($1300 < R < 2000$) visible and infrared spectra. We select a sample of 112 wide binary systems that contain a late-type companion to a solar-type primary star. Our sample includes 62 primary stars with previously published metallicities, as well as 50 stars with metallicities determined from our own observations. We use our sample to empirically determine which features in the spectrum of the companion are best correlated with the metallicity of the primary. We find $\simeq$ 120 features in K and M dwarf spectra that are useful for predicting metallicity. We derive metallicity calibrations for different wavelength ranges, and show that it is possible to get metallicities reliable to $\lesssim 0.10$~dex using either visible, $J$-, $H$-, or $K$- band spectra. We find that the most accurate metallicities derived from visible spectra requires the use of different calibrations for early-type (K5.5--M2) and late-type (M2--M6) dwarfs. Our calibrations are applicable to dwarfs with metallicities of $-1.04 < $[Fe/H]$ < +0.56$ and spectral types from K7 to M5. Lastly, we use our sample of wide binaries to test and refine existing calibrations to determine M dwarf metallicities. We find that the $\zeta$ parameter, which measures the ratio of TiO can CaH bands, is correlated with [Fe/H] for super-solar metallicities, and $\zeta$ does not always correctly identify metal-poor M dwarfs. We also find that existing calibrations in the $K$ and $H$ bands are quite reliable for stars with [Fe/H]$>-0.5$, but are less useful for more metal-poor stars.
\end{abstract}

\keywords{binaries: visual -- stars: abundances, -- , stars: fundamental parameters, -- stars: late-type, -- techniques: spectroscopic}

\section{Introduction}\label{sec:intro}
Despite their intrinsic faintness, M dwarfs have become attractive targets for exoplanet searches \citep[e.g.,][]{Charbonneau:2009rt, Apps:2010zr, Mann:2011qy}, as M dwarfs have smaller radii and lower masses than their solar-type counterparts, allowing for easier detection of low-mass exoplanets \citep{Gaidos:2007ly}. For transiting M dwarf planets with high-precision observations, such as those done by {\it Kepler} \citep{Borucki:2010lr, Batalha:2013lr}, errors in planet parameters (primarily planet radius) are directly tied to errors in stellar parameters \citep[e.g.,][]{Muirhead:2012pd}, which in turn depend on reliable measurements of stellar temperature and metallicity \citep{Dotter:2008fk, Demory:2009qy, Allard:2011lr}. Accurate metallicities are also necessary to study any correlation between metallicity and planet frequency \citep[e.g.][]{Johnson:2010lr, Mann:2012}. Further, because the stellar mass function peaks around mid-M, M dwarfs weigh heavily in any study of Galactic structure. The distribution of M dwarf metallicities can therefore be used to set limits on Milky Way formation models \citep{Woolf:2012lr}. However, unlike those of their solar-type counterparts, M-dwarf metallicities are difficult to determine, primarily due to the presence of complex molecular lines in their visible spectra, which result in line confusion and a lack of identifiable continuum, and do not always match with current M dwarf models \citep{1976A&A....48..443M, Allard:2011lr}. The visible wavelengths usually used to derive metallicities for solar-type stars are dominated by TiO lines that increase in strength with decreasing effective temperature. These TiO lines obscure the continuum, making equivalent width measurements unreliable, and differ as a function of spectral type. Direct spectral synthesis has been tried on a small sample of stars \citep{Bean:2006fk,Onehag:2012lr}. However, such spectra is observationally expensive, and spectral synthesis is complicated by incomplete lines lists.
 
One common technique is to use wide binaries with a solar-type (late-F, G, or early-K) primary and a late-type (late K or M) dwarf companion. Since these stars presumably formed from the same molecular cloud, the metallicity of the M dwarf companion can be assumed to be the same as that of the FGK primary \citep{2005A&A...442..635B}. From this one can create empirical calibrations of observable features to determine metallicities in M dwarfs. Techniques based on absolute photometry \citep{2005A&A...442..635B,Johnson:2009fk, Schlaufman:2010qy} require parallaxes, which are only available for a handful of the closest and brightest M dwarfs. Molecular indices at visible wavelengths \citep[e.g.][]{Woolf:2006uq, Lepine:2013lr} have proven useful for separating M dwarfs into luminosity/metallicity classes (i.e., dwarf, subdwarf, and extreme subdwarf), but saturate near solar metallicity and are less reliable for late-K and early-M dwarfs. 

The use of spectral indices in the infrared has, however, been showing considerable promise. \citet[][henceforth R10]{2010ApJ...720L.113R} showed that [Fe/H] can be inferred for M dwarfs using the Na~I doublet and Ca~I triplet in moderate resolution ($R\simeq$2700) $K$-band spectra and assuming solar relative abundances. \citet[][henceforth R12)]{Rojas-Ayala:2012uq} expanded on this, showing that one can determine [M/H] for early to mid-M dwarfs as accurately as 0.1~dex using the same indices. Further, \citet[][henceforth T12]{Terrien:2012lr} were able to demonstrate similar precision in the $H-$band. However, calibrations for both $H$- and $K$-band metallicities were derived using a relatively small sample (18 and 22, respectively) of wide binaries. As a result, they were only verified for systems with near-solar metallicities ($-0.4 \lesssim$ [Fe/H] $\lesssim$ +0.3), and for a narrow range of spectral types (M0--M4). 

In this paper we analyze visible and near-infrared (NIR) observations of 112 late-K and M dwarfs with F, G, or early K star primaries. We provide determine metallicity estimates for 50 of the FGK stars. We also provide a full list of the $120$ features we identify to be metal-sensitive features in visible and NIR spectra, with the expectation that these will be useful for future studies on M dwarf metallicities. In Section~\ref{sec:sample} we present our sample of wide-binary systems, followed by a description of our observations and reduction in Section~\ref{sec:obs}. In Section~\ref{sec:analysis} we explain how we determine basic properties for both the primary and companion stars. We discuss our search for metal-sensitive features in Section~\ref{sec:MCanalysis} and present the results from this analysis in Section~\ref{sec:determinemetal}, which includes our calibrations to determine metallicities at a range of wavelengths. We test existing techniques in Section~\ref{sec:recal}. Lastly, we summarize the results in Section 8 and discuss possible drawbacks of our analysis as well as prospects for future studies. All wavelengths used in this work are stated as vacuum values.

\section{Sample}\label{sec:sample}
We draw a sample of late K and M dwarfs with comoving FGK stars from a variety of literature sources, specifically \citet{Chaname:2004lr}, \citet{Gould:2004fk}, \citet{Lepine:2007qy}, \citet{Schlaufman:2010qy}, and \citet{Tokovinin:2012fj}. We also identify a number of new common-proper motion (CPM) pairs that contain an FGK star and a late K or M star using proper motions from (in order of preference) {\it Hipparcos} \citep{van-Leeuwen:2005kx,van-Leeuwen:2007yq}, SUPERBLINK \citep{2005AJ....129.1483L}, or the PPMXL survey \citep{2010AJ....139.2440R}. We identify new CPM pairs, as well as vet the literature sample, following the techniques of \citet{Lepine:2007qy} and \citet{Dhital:2010lr}. \citet{Lepine:2007qy} identify CPM pairs using the formula:
\begin{equation}
\Delta X = [(\mu/0.15)^{-3.8}\Delta \theta \Delta \mu]^{0.5},
\end{equation}
where $\mu$ is the absolute proper motion of the primary star in arcsec~yr$^{-1}$, $\Delta \theta$ is the separation between the two stars in arcsec, and $\Delta \mu$ is the absolute difference in proper motion between the two stars in arcsec~yr$^{-1}$. \citet{Lepine:2007qy} consider two stars to be physically associated with each other if $\Delta X<1$, $\Delta \theta<1500\arcsec$, and $\Delta \mu < 100$~mas~yr$^{-1}$. We take a more conservative cut and require that our pairs have $\Delta X<0.9$, $\Delta \theta<650\arcsec$, and $\Delta \mu < 60$~mas~yr$^{-1}$ to cut down on the number of chance alignments. This technique takes advantage of the absolute proper motion, and not just the difference in proper motion, of the star, however, it does not factor in associated errors in proper motion. Thus we further force the constraint from \citet{Dhital:2010lr}: 
\begin{equation}\label{eqn:properr}
\left( \frac{\Delta\mu_\alpha}{\sigma_{\Delta \mu_{\alpha}}}\right)^2 + \left( \frac{\Delta\mu_\delta}{\sigma_{\Delta \mu_{\delta}}}\right)^2 \le 2,
\end{equation}
where $\Delta\mu_\alpha$ and $\Delta\mu_\delta$ are the differences in proper motion between the two components (right ascension and declination, respectively) and $\sigma_{\Delta \mu}$ is the error in the proper motion differences. Equation~\ref{eqn:properr} is designed to ensure that the resulting sample has minimal contamination from chance alignment pairs, but at the cost of excluding a large number of true wide binaries. To increase our sample, we add pairs that do not satisfy Equation 2 but have published parallaxes for both the primary and companion star consistent to 1$\sigma$ or the parallax uncertainty, which strongly suggests that the pair forms a physical system

We remove pairs with $\delta<-45^\circ$ or $\delta>+68^\circ$, since these are outside the reach of telescopes in Hawaii, and pairs with $\Delta \theta<3\arcsec$ as these will be difficult to observe and may have inaccurate photometry. We further remove pairs with $V_{c}-J_{c}<2.5$ (the subscript $c$ denoting the companion and $p$ the primary). This cut will remove almost all stars earlier than K5 \citep{Lepine:2011vn} where metallicities can be measured using modified spectral synthesis techniques \citep{1996A&AS..118..595V, 2005ApJS..159..141V}. We cut out systems with $V_{c}>18$, and $K_{c}>12$, as the observation time required for these stars is highly prohibitive. For the same reason, we remove systems with $V_{p}>12$ unless the primary star already has a published metallicity. The resulting sample contains 262 pairs with primaries with colors in the range $0.8<V_p-J_p<2.5$ and/or published temperatures consistent with a late-F, G, or early-K dwarf \citep{1970A&A.....4..234F, 2001ApJ...558..309D}. From here we prioritize our observations based on; (1) $V_{c}$ and $K_{c}$ magnitudes (to minimize required telescope time), (2) $V_{c}-J_{c}$ color (to ensure a range of spectral types), (3) availability of metallicities for the primary star in the literature or $V_{p}$ magnitude if no literature metallicity is available (see Section~\ref{sec:cfhtobs}), (4) availability of parallax information for the primary or secondary, which makes it more likely that these are true CPM pairs, and (5) metallicity of the primary (if available) to guarantee a range of metallicities for our analysis. 

The resulting sample includes 112 stars for which we have obtained infrared spectra (Section~\ref{sec:spex}), visible wavelength spectra (Section~\ref{sec:snifs}), and metallicities for the primary star from the literature or from our own analysis (Section~\ref{sec:FGK}). Our sample has $-1.04<$[Fe/H]$_p<0.56$, and companion star spectral types from K5.5 to M6. We list our sample in Table \ref{tab:sample}.

\begin{deluxetable*}{l l l l | l l l l r l l r l}
\caption{Wide Binary Sample}
\tablewidth{0pt}
\tablehead{
\multicolumn{3}{c}{Companion} & \multicolumn{7}{c}{Primary} \\
\cline{1-4}  \cline{5-11} \\ 
\colhead{Common Name} & \colhead{R.A}& \colhead{$\delta$} & \colhead{SpT$^a$} & \colhead{Name} & \colhead{[Fe/H]$^b$} & \colhead{$\pm$} & \colhead{Source$^c$} & \colhead{[M/H]$^b$} & \colhead{$\pm$} & \colhead{Source$^c$} \\
}
\startdata
NLTT 738 & 3.8107 & $+53.0794$ & M2.2 & HIP 1224 & $+0.07$ & 0.03 & This work & $+0.03$ & 0.03 & This work\\
NLTT 923 & 4.6075 & $+44.0272$ & M4.1 & HIP 1475 & $-0.05$ & 0.03 & VF05 & $-0.53$ & 0.03 & VF05\\
NLTT 2478 & 11.3066 & $+0.2642$ & M3.8 & HIP 3540 & $+0.02$ & 0.03 & VF05 & $-0.01$ & 0.03 & VF05\\
NLTT 3598 & 16.3740 & $+15.3885$ & K7.4 & HIP 5110 & $-0.09$ & 0.03 & This work & $-0.13$ & 0.03 & This work\\
NLTT 3725 & 16.9105 & $+22.9558$ & M3.9 & HIP 5286 & $+0.25$ & 0.03 & This work & $+0.21$ & 0.05 & This work\\
Gl 56.3 & 19.6676 & $-0.8744$ & K7.6 & HIP 6130 & $-0.01$ & 0.06 & C01 & $+0.01$ & 0.06 & C11\\
NLTT 4568 & 20.6531 & $+12.7524$ & K7.8 & HIP 6431 & $+0.10$ & 0.03 & This work & $+0.04$ & 0.03 & This work\\
NLTT 4599 & 20.7525 & $-12.9583$ & K7.9 & HIP 6456 & $+0.45$ & 0.03 & This work & $+0.36$ & 0.06 & This work\\
Gl 81.1 & 29.2961 & $-10.2481$ & M0.1 & HIP 9094 & $+0.12$ & 0.03 & VF05 & $+0.05$ & 0.03 & VF05\\
Gl 100 & 37.1329 & $-20.0407$ & M2.9 & HIP 11565 & $-0.28$ & 0.04 & N12 & ... & ... & ...\\
NLTT 8107 & 37.2792 & $+22.8671$ & K5.9 & HIP 11572 & $-0.08$ & 0.03 & This work & $+0.00$ & 0.03 & This work\\
Gl 105 & 39.0636 & $+6.8717$ & M3.9 & HIP 12114 & $-0.12$ & 0.03 & VF05 & $-0.00$ & 0.03 & VF05\\
NLTT 8787 & 41.0427 & $+49.2317$ & M1.6 & HIP 12777 & $+0.06$ & 0.03 & VF05 & $+0.02$ & 0.03 & VF05\\
Gl 118.2 & 43.8991 & $+26.8724$ & M3.8 & HIP 13642 & $+0.28$ & 0.03 & VF05 & $+0.21$ & 0.03 & VF05\\
NN 3195 & 46.1810 & $+61.7358$ & M2.9 & HIP 14286 & $-0.28$ & 0.03 & S11 & $-0.26$ & 0.06 & C11\\
NLTT 10349 & 48.7538 & $+1.0523$ & M1.0 & HIP 15126 & $-0.92$ & 0.03 & This work & $-0.66$ & 0.03 & This work\\
NLTT 11125 & 53.0251 & $+43.6669$ & K7.3 & HIP 16467 & $-0.01$ & 0.03 & This work & $-0.01$ & 0.03 & This work\\
NLTT 11176 & 53.3085 & $+46.2553$ & K7.3 & HIP 16563 & $+0.20$ & 0.03 & This work & $+0.16$ & 0.06 & This work\\
NLTT 11270 & 53.8689 & $+42.8930$ & M0.4 & NLTT 11280 & $-0.40$ & 0.08 & C11 & $-0.30$ & 0.06 & C11\\
NLTT 11500 & 54.8955 & $+18.3143$ & M1.8 & HIP 17076 & $-0.52$ & 0.08 & C11 & $-0.44$ & 0.06 & C11\\
PM J0355+5214 & 58.9037 & $+52.2414$ & M2.7 & HIP 18366 & $-0.36$ & 0.05 & Ra07 & $-0.33$ & 0.06 & C11
\enddata
\tablecomments{Table \ref{tab:sample} is published in its entirety in the electronic edition of the {\it Astronomical Journal}, and can be downloaded with the arXiv version of the manuscript. A portion is shown here for guidance regarding its form and content.}
\tablenotetext{a}{Spectral types derived from TiO and CaH indices, (see \citet{Lepine:2013lr}). Continuous spectral types (to 0.1) are used for plotting/binning/calculations, even though spectral types are only accurate to $\pm$0.2 (and by convention should be rounded to the nearest 0.5).}
\tablenotetext{b}{[M/H] and [Fe/H] values shown here include our applied corrections (see Section~\ref{sec:analysis}).}
\tablenotetext{c}{Metallicity sources: 
C01 = \citet{2001A&A...373..159C},
M04 = \citet{2004A&A...418..551M}, 
LH05 = \citet{2005AJ....129.1063L}, 
VF05 = \citet{2005ApJS..159..141V},
T05 = \citet{2005PASJ...57...27T}, 
B06 = \citet{Bean:2006fk},
Ra07 = \citet{2007A&A...465..271R},
Ro07 = \citet{2007ApJS..169..430R},
F08 = \citet{2008MNRAS.384..173F}, 
S11 = \citet{2011A&A...526A..71D}, 
C11 = \citet{2011A&A...530A.138C}, 
N12 = \citet{2012A&A...538A..25N},
This work = analysis of CFHT/ESPaDOnS spectra as part of this program.\\
Note that all metallicities sources are from high-resolution spectra, with the exception of Ro07, which uses moderate-resolution spectra, and C11, which uses Str\"{o}mgren photometry.
}
\label{tab:sample}
\end{deluxetable*}

\section{Observations and Reduction}\label{sec:obs}
\subsection{ESPaDOnS/CFHT}\label{sec:cfhtobs}
Between 2011 January and 2012 April, 60 F-, G- and early K-type stars were observed using the ESPaDOnS spectrograph attached to the Canada France Hawaii Telescope \citep[CFHT; ][]{Donati:2003uq} on Mauna Kea. Observations were taken in the star+sky mode, which gave a resolution of R$\simeq 65000$ and a wavelength range from 0.37\um\ to 1.05\um. All observations were designed to achieve a signal-to-noise ratio (S/N) of $>100$ at 0.67\um, and typical S/N was $>150$ (per resolving element). The data were reduced automatically using the {\it Libre-ESpRIT} pipeline described in \citet{Donati:1997fj}. 

\subsection{SpeX/IRTF}\label{sec:spex}
We obtained near-infrared spectra of our sample of companions using the SpeX spectrograph \citep{Rayner:2003lr} attached to the NASA Infrared Telescope Facility (IRTF) on Mauna Kea. SpeX observations were taken in the short cross-dispersed (SXD) mode using the 0.3$\times15\arcsec$ slit, yielding simultaneous coverage from 0.8 to 2.4\um\ and a resolution of $R\simeq2000$. The star was placed at two positions along the slit (A and B). Exposures were taken with an ABBA slit-nodding pattern, with at least 6 exposures in total. Integration times were no longer than 120~s for each exposure to minimize the effect of changes in atmospheric H$_2$O. Thus for faint stars more than 6 exposures were required to get sufficient S/N. S/N in the $H$- and $K$-bands was $>100$ in all cases, and typically $>150$ (per resolving element). To correct for telluric lines, we observed an A0V-type star within 30 minutes and 0.1 air mass of the target observation (and usually much closer in time and air mass). Often the same A0V star was used to remove telluric lines for more than one target. To remove effects from large telescope slews, we obtained flat-field and argon lamp calibration sequences after each A0V star.

Spectra were extracted and reduced using the SpeXTool package \citep{Cushing:2004fk}, which performed flat-field correction, wavelength calibration, sky subtraction, and extraction of the one-dimentional (1D) spectrum. Multiple exposures were stacked using the IDL routine {\it xcombxpec}. A telluric correction spectrum was constructed from each A0V star using the {\it xtellcor} package \citep{Vacca:2003qy}, and then applied to the relevant target spectra. 

Reduced spectra were put in vacuum wavelengths using the formula from \citet{Ciddor:96}. We put spectra in the star's rest frame by comparing them to a spectrum of the template star HD36395 (an M1.5 dwarf, also in rest frame/vacuum) taken from the IRTF spectral library \citep{Cushing:2005lr, Rayner:2009kx}. We cross-correlated each spectrum with the template, in orders 3 -- 7 separately (order 8 is ignored because it is too smooth and has relatively poor S/N), yielding 6 radial velocities (RVs). We shifted the spectrum by the average (after removing any $5\sigma$ outliers) of each set of RVs. 

\subsection{SNIFS/UH2.2m}\label{sec:snifs}
We obtained a visible spectrum of each companion with the SuperNova Integral Field Spectrograph \citep[SNIFS,][]{lantz:2004} on the University of Hawaii 2.2m telescope atop Mauna Kea. SNIFS has $R\simeq1300$ and splits the signal with a dichroic mirror into blue (0.32--0.52\,\um) and red (0.52--0.95\,\um) channels. SNIFS data processing is performed with a data reduction pipeline, described in detail in \citet{Bacon:2001} and \citet{Aldering:2006}. SNIFS processing includes dark, bias, and flat-field corrections, assembling the data into red and blue three-dimentional data cubes, and cleaning them for cosmic rays and bad pixels. Wavelengths are calibrated with arc lamp exposures taken at the same telescope pointing as the science data. The calibrated spectrum is then sky-subtracted, and a 1-D spectrum is extracted using a point-spread function model. Corrections are applied to the spectrum for instrument response, and for telluric lines based on observations of the Feige 66, Feige 110, BD+284211, or BD+174708 spectrophotometric standards \citep{Oke:1990} that are taken over the course of each night.

Approximate spectral types are determined by the HAMMER software package \citep{2007AJ....134.2398C}. The spectra are then shifted to zero radial velocity by cross-correlating with templates from \citet{Bochanski:2007lr} of the corresponding spectral type. Late-K stars are cross-correlated using an M0 template. 

\subsection{Construction of a combined M dwarf Spectrum}
We use the overlapping region in our SpeX and SNIFS data (0.81--0.96\um) to combine the visible and NIR spectra. We normalize each SNIFS spectrum by a constant, $C$, which is equal to the ratio of the median flux of the SNIFS spectra in overlapping region to the median flux of the SpeX spectra in the overlapping region. We also find that there is a systematic offset in wavelength between the visible wavelength and NIR spectra in the overlapping region. This amounts to a RV shift of $\simeq 30$~km~s$^{-1}$ between the SNIFS and SpeX spectra. The most likely explanation for this is a small difference in the RV templates used for our SpeX and SNIFS data, taken from \citet{Rayner:2009kx} and \citet{Bochanski:2007lr} respectively. We choose to shift the visible wavelength data to match with the NIR data (which has higher resolution and therefore gives more reliable RVs) by adding an additional RV correction of $-30$~km~s$^-1$ to each SNIFS spectrum. Given the modest resolution of SNIFS data, this correction is unlikely to significantly change our results. The offset corresponds to $<1$\,\AA, which is significantly less than the resolving power of SNIFS, and is similar in size to random errors in our RV measurements. 

\section{Deriving Stellar Parameters}\label{sec:analysis}
\subsection{FGK Metallicities}\label{sec:FGK}
We draw primary star metallicities from a variety of sources in the literature. We list the adopted metallicity and literature source for each binary in Table~\ref{tab:sample}. In total, 33 primary star metallicities come from the SPOCS catalog \citep{2005ApJS..159..141V}, 50 from observations taken as part of this project with CFHT/ESPaDOnS, and 29 from other literature sources. SPOCS consists of high-resolution echelle spectra of $>1000$ F-, G-, and K-type stars obtained with the Keck, Lick, or Anglo-Australian Telescope. \citet{2005ApJS..159..141V} fit the observed spectrum to a synthetic spectrum using the software package SME \citep[Spectroscopy Made Easy,][]{1996A&AS..118..595V}, which provides a set of observational parameters (\teff, [Fe/H], [M/H], log~$g$, etc.) for each star. We adopt an uncertainty of 0.03 dex for their derived [M/H] and [Fe/H] values. 

To determine the stellar parameters of primaries observed with CFHT/ESPaDOnS we model each spectrum using the SME software \citep{1996A&AS..118..595V}, fitting the spectrum to a set of tuned lines from the SPOCS catalog \citep{2005ApJS..159..141V}.  We simultaneously solve for surface gravity, effective temperature, projected rotational velocity, and metallicity in addition to individual abundances of Fe, Na, Si, Ni, and Ti.  Solar values are assumed for all of the initial models and after obtaining an initial fit, we then perturb \teff by $\pm 100$K and fit again.  Our final model parameters are $\chi^2$-weighted averages of three runs.  Corrections based on Vesta and stellar binary observations as detailed in \citet{2005ApJS..159..141V} are then applied. 

For stars with good parallax measurements ({\it Hipparcos} stars or their companions), we use Yonsei-Yale (Y$^2$) isochrones \citep{2004ApJS..155..667D} to better constrain the surface gravity \citep{Valenti:2009fk}.  After we determine the stellar parameters as above, we use distance and $B$ and $V$ magnitudes to the derive bolometric luminosity and, combined with the SME \teff, the stellar radius. Bolometric corrections are obtained by interpolating in the high temperature grid of \citet{VandenBerg:2003lr}, and $B$, $V$ magnitudes were drawn from {\it Hipparcos} \citep{van-Leeuwen:2005kx}. The SME determined ratio of Si to Fe is used as a proxy for alpha element enhancement.  A best-fit evolutionary model is found by interpolating in the Y$^2$ grid which yielded a surface gravity for the star.  This log~$g$ is compared to the value determined using SME and if the two did not match, a new set of SME models is found with the gravity fixed to the isochrone value. The process is repeated until the log~$g$ values agree to within 0.001 dex. Final stellar parameters (\teff, log~$g$, [M/H], etc.) for stars observed as part of our program are listed in Table~\ref{tab:primaries}.

\begin{table*} 
\setlength{\tabcolsep}{0.04in}
\caption{Parameters of Primary Stars observed at CFHT}
\begin{centering}
\begin{tabular}{l  cccccccccccccccccll}
\hline \hline
Name & \teff & $\pm$ & log~$g$ & $\pm$ & [Fe/H] & $\pm$ & [M/H]  & $\pm$ & [Na/H] & $\pm$ & [Ti/H] & $\pm$ & [Si/H] & $\pm$ & [Ni/H] & $\pm$ & $\chi_{\mathrm{red}}^2$ & Run Type$^a$ \\  \hline
HIP 1224 & 5141 &  44 &  4.53 &  0.06 & $+0.07$ &  0.03 & $+0.03$ &  0.03 & $+0.12$ &  0.03 & $+0.03$ &  0.05 & $+0.02$ &  0.02 & $+0.08$ &  0.03 &  2.9 & ITER \\
HIP 5110 & 4648 &  44 &  4.61 &  0.06 & $-0.09$ &  0.03 & $-0.13$ &  0.03 & $-0.25$ &  0.03 & $-0.18$ &  0.05 & $-0.15$ &  0.02 & $-0.11$ &  0.03 &  6.0 & ITER \\
HIP 5286 & 4676 &  53 &  4.60 &  0.06 & $+0.25$ &  0.03 & $+0.21$ &  0.05 & $+0.30$ &  0.09 & $+0.14$ &  0.05 & $+0.09$ &  0.06 & $+0.26$ &  0.03 & 14.7 & ITER \\
HIP 6431 & 4858 &  44 &  4.57 &  0.06 & $+0.10$ &  0.03 & $+0.04$ &  0.03 & $+0.10$ &  0.03 & $+0.00$ &  0.05 & $+0.06$ &  0.02 & $+0.11$ &  0.03 &  6.3 & ITER \\
HIP 6456 & 5222 &  52 &  4.44 &  0.06 & $+0.45$ &  0.03 & $+0.36$ &  0.06 & $+0.62$ &  0.05 & $+0.27$ &  0.08 & $+0.43$ &  0.02 & $+0.53$ &  0.03 &  6.8 & ITER \\
HIP 11572 & 5093 &  44 &  4.51 &  0.06 & $-0.08$ &  0.03 & $+0.00$ &  0.03 & $+0.05$ &  0.03 & $+0.03$ &  0.05 & $+0.05$ &  0.02 & $-0.02$ &  0.03 &  3.6 & ITER \\
HIP 15126 & 5285 &  44 &  4.64 &  0.06 & $-0.92$ &  0.03 & $-0.66$ &  0.03 & $-0.84$ &  0.03 & $-0.56$ &  0.05 & $-0.51$ &  0.02 & $-0.86$ &  0.03 &  2.2 & ITER \\
HIP 16467 & 5539 &  44 &  4.33 &  0.06 & $-0.01$ &  0.03 & $-0.01$ &  0.03 & $-0.09$ &  0.03 & $-0.04$ &  0.05 & $+0.02$ &  0.02 & $-0.04$ &  0.03 &  2.1 & ITER \\
HIP 16563 & 5788 &  44 &  4.52 &  0.06 & $+0.20$ &  0.03 & $+0.16$ &  0.06 & $+0.10$ &  0.03 & $+0.21$ &  0.05 & $+0.04$ &  0.02 & $+0.05$ &  0.04 &  4.4 & ITER \\
HIP 31127 & 5158 &  44 &  3.83 &  0.06 & $-0.54$ &  0.03 & $-0.45$ &  0.03 & $-0.56$ &  0.03 & $-0.50$ &  0.05 & $-0.34$ &  0.02 & $-0.56$ &  0.03 &  3.1 & ITER \\
HIP 31597 & 5428 &  44 &  4.44 &  0.06 & $+0.09$ &  0.03 & $+0.08$ &  0.03 & $+0.14$ &  0.03 & $+0.08$ &  0.05 & $+0.10$ &  0.02 & $+0.12$ &  0.03 &  2.0 & ITER \\
HIP 32423 & 4817 &  44 &  4.64 &  0.06 & $-0.26$ &  0.03 & $-0.21$ &  0.04 & $-0.32$ &  0.04 & $-0.27$ &  0.07 & $-0.24$ &  0.03 & $-0.27$ &  0.03 &  5.6 & ITER \\
HIP 35449 & 6156 &  44 &  4.36 &  0.06 & $+0.21$ &  0.03 & $+0.18$ &  0.03 & $+0.17$ &  0.03 & $+0.21$ &  0.05 & $+0.19$ &  0.02 & $+0.18$ &  0.03 &  1.9 & ITER \\
HIP 40298 & 5618 &  44 &  4.51 &  0.06 & $-0.07$ &  0.03 & $-0.09$ &  0.03 & $-0.09$ &  0.03 & $-0.04$ &  0.05 & $-0.08$ &  0.02 & $-0.14$ &  0.03 &  1.7 & ITER \\
NLTT 12373 & 5838 &  93 &  4.62 &  0.09 & $-0.08$ &  0.05 & $-0.00$ &  0.08 & $-0.08$ &  0.05 & $+0.05$ &  0.08 & $-0.04$ &  0.03 & $-0.12$ &  0.06 &  2.2 & VESTA \\
HIP 45863 & 5172 &  44 &  4.51 &  0.06 & $-0.12$ &  0.03 & $-0.13$ &  0.03 & $-0.07$ &  0.03 & $-0.10$ &  0.05 & $-0.09$ &  0.02 & $-0.12$ &  0.03 &  4.3 & ITER \\
NLTT 23002 & 5259 &  44 &  4.58 &  0.06 & $-0.15$ &  0.03 & $-0.11$ &  0.03 & $-0.12$ &  0.03 & $-0.11$ &  0.05 & $-0.14$ &  0.02 & $-0.16$ &  0.03 &  3.1 & VESTA \\
HIP 50802 & 4472 &  44 &  4.70 &  0.06 & $-0.01$ &  0.03 & $-0.06$ &  0.03 & $-0.25$ &  0.04 & $-0.11$ &  0.05 & $-0.21$ &  0.06 & $-0.14$ &  0.03 & 13.1 & ITER \\
HIP 54155 & 5547 &  50 &  4.54 &  0.06 & $+0.16$ &  0.03 & $+0.11$ &  0.07 & $+0.03$ &  0.03 & $+0.14$ &  0.06 & $+0.05$ &  0.02 & $+0.05$ &  0.03 &  2.9 & ITER \\
HIP 55486 & 5371 &  44 &  4.51 &  0.06 & $+0.46$ &  0.03 & $+0.42$ &  0.03 & $+0.58$ &  0.07 & $+0.37$ &  0.05 & $+0.39$ &  0.02 & $+0.48$ &  0.03 &  9.6 & ITER \\
HIP 56729 & 5421 &  44 &  4.47 &  0.06 & $-0.09$ &  0.03 & $-0.04$ &  0.03 & $-0.10$ &  0.03 & $-0.02$ &  0.05 & $-0.04$ &  0.02 & $-0.11$ &  0.03 &  2.7 & ITER \\
HIP 56930 & 5025 &  44 &  4.61 &  0.06 & $-0.12$ &  0.03 & $-0.17$ &  0.05 & $-0.20$ &  0.03 & $-0.15$ &  0.05 & $-0.14$ &  0.02 & $-0.18$ &  0.03 &  7.0 & ITER \\
HIP 59080 & 5423 &  44 &  4.41 &  0.06 & $-0.16$ &  0.03 & $-0.13$ &  0.03 & $-0.12$ &  0.03 & $-0.16$ &  0.05 & $-0.11$ &  0.02 & $-0.16$ &  0.03 &  4.9 & ITER \\
HIP 61081 & 5298 &  44 &  4.59 &  0.06 & $-0.54$ &  0.03 & $-0.42$ &  0.03 & $-0.49$ &  0.03 & $-0.33$ &  0.05 & $-0.26$ &  0.02 & $-0.48$ &  0.03 &  2.5 & ITER \\
HIP 61189 & 4655 &  44 &  4.93 &  0.06 & $+0.11$ &  0.03 & $+0.05$ &  0.03 & $-0.01$ &  0.03 & $+0.02$ &  0.05 & $+0.08$ &  0.02 & $+0.09$ &  0.03 & 16.6 & VESTA \\
HIP 61589 & 5700 &  44 &  4.49 &  0.06 & $-0.05$ &  0.03 & $-0.08$ &  0.05 & `$-0.08$ &  0.03 & $-0.05$ &  0.05 & $-0.05$ &  0.02 & $-0.06$ &  0.03 &  2.3 & ITER \\
HIP 64345 & 5495 &  44 &  4.40 &  0.06 & $-0.57$ &  0.03 & $-0.39$ &  0.03 & $-0.49$ &  0.03 & $-0.33$ &  0.05 & $-0.25$ &  0.02 & $-0.52$ &  0.03 &  1.8 & ITER \\
HIP 64797 & 5041 &  44 &  4.60 &  0.06 & $-0.12$ &  0.03 & $-0.13$ &  0.03 & $-0.16$ &  0.03 & $-0.11$ &  0.05 & $-0.16$ &  0.02 & $-0.17$ &  0.03 & 15.1 & ITER \\
HIP 65636 & 4619 &  44 &  4.64 &  0.06 & $+0.13$ &  0.03 & $+0.01$ &  0.03 & $-0.11$ &  0.03 & $-0.08$ &  0.05 & $+0.00$ &  0.02 & $+0.02$ &  0.03 & 13.4 & ITER \\
HIP 65963 & 5478 &  44 &  4.52 &  0.06 & $-0.14$ &  0.03 & $-0.08$ &  0.03 & $-0.19$ &  0.03 & $-0.08$ &  0.05 & $-0.09$ &  0.02 & $-0.14$ &  0.03 &  1.7 & ITER \\
HIP 68799 & 5492 &  44 &  4.42 &  0.06 & $-0.03$ &  0.03 & $-0.06$ &  0.03 & $+0.00$ &  0.03 & $-0.05$ &  0.05 & $-0.02$ &  0.02 & $-0.04$ &  0.03 &  2.7 & ITER \\
HIP 70100 & 4886 &  44 &  4.57 &  0.06 & $+0.15$ &  0.03 & $+0.13$ &  0.03 & $+0.26$ &  0.03 & $+0.09$ &  0.05 & $+0.12$ &  0.02 & $+0.16$ &  0.03 &  9.9 & ITER \\
HIP 70426 & 4801 &  44 &  4.58 &  0.06 & $+0.09$ &  0.03 & $+0.03$ &  0.04 & $+0.13$ &  0.04 & $-0.00$ &  0.05 & $+0.06$ &  0.02 & $+0.10$ &  0.03 & 12.8 & ITER \\
HIP 74396 & 5124 &  44 &  4.60 &  0.06 & $-0.09$ &  0.03 & $-0.07$ &  0.03 & $-0.08$ &  0.03 & $-0.06$ &  0.05 & $-0.12$ &  0.02 & $-0.13$ &  0.03 &  3.5 & ITER \\
HIP 74734 & 5822 &  44 &  4.34 &  0.06 & $-0.32$ &  0.03 & $-0.27$ &  0.03 & $-0.37$ &  0.03 & $-0.20$ &  0.05 & $-0.23$ &  0.02 & $-0.39$ &  0.03 &  2.1 & ITER \\
HIP 75069 & 5196 &  44 &  4.61 &  0.06 & $-0.38$ &  0.03 & $-0.34$ &  0.03 & $-0.36$ &  0.03 & $-0.33$ &  0.05 & $-0.33$ &  0.02 & $-0.40$ &  0.03 &  2.8 & ITER \\
HIP 76668 & 4636 &  44 &  4.65 &  0.06 & $-0.06$ &  0.03 & $-0.14$ &  0.03 & $-0.22$ &  0.06 & $-0.17$ &  0.05 & $-0.10$ &  0.02 & $-0.12$ &  0.03 & 13.7 & ITER \\
HIP 78969 & 4899 &  44 &  4.58 &  0.06 & $+0.19$ &  0.03 & $+0.12$ &  0.03 & $+0.23$ &  0.06 & $+0.08$ &  0.05 & $+0.13$ &  0.02 & $+0.20$ &  0.03 &  9.0 & ITER \\
HIP 79629 & 5600 &  50 &  4.48 &  0.06 & $-0.25$ &  0.04 & $-0.25$ &  0.06 & $-0.28$ &  0.03 & $-0.25$ &  0.05 & $-0.20$ &  0.02 & $-0.31$ &  0.03 &  2.0 & ITER \\
HIP 84616 & 4826 &  44 &  4.62 &  0.06 & $-0.12$ &  0.03 & $-0.09$ &  0.03 & $-0.09$ &  0.03 & $-0.09$ &  0.05 & $-0.10$ &  0.02 & $-0.14$ &  0.03 &  6.4 & ITER \\
PM J1742+1645 & 5492 &  44 &  4.50 &  0.06 & $-0.09$ &  0.03 & $+0.01$ &  0.03 & $-0.06$ &  0.03 & $-0.00$ &  0.05 & $-0.11$ &  0.02 & $-0.15$ &  0.03 &  4.2 & VESTA \\
HIP 87082 & 5795 &  44 &  4.40 &  0.06 & $-0.05$ &  0.03 & $-0.03$ &  0.05 & $-0.17$ &  0.03 & $+0.02$ &  0.05 & $-0.02$ &  0.02 & $-0.09$ &  0.03 &  2.7 & ITER \\
HIP 88188 & 5299 &  53 &  4.54 &  0.06 & $+0.05$ &  0.04 & $-0.04$ &  0.07 & $-0.06$ &  0.03 & $-0.05$ &  0.13 & $+0.02$ &  0.03 & $+0.04$ &  0.03 &  3.2 & ITER \\
HIP 88365 & 5371 &  44 &  4.59 &  0.06 & $-0.66$ &  0.03 & $-0.34$ &  0.03 & $-0.52$ &  0.03 & $-0.33$ &  0.05 & $-0.33$ &  0.02 & $-0.57$ &  0.03 &  3.7 & ITER \\
HIP 90246 & 4494 &  44 &  4.69 &  0.06 & $-0.01$ &  0.03 & $-0.08$ &  0.03 & $-0.26$ &  0.05 & $-0.08$ &  0.05 & $-0.08$ &  0.05 & $-0.09$ &  0.03 & 11.3 & ITER \\
HIP 104097 & 4428 &  44 &  4.72 &  0.06 & $-0.38$ &  0.03 & $-0.27$ &  0.03 & $-0.51$ &  0.03 & $-0.15$ &  0.05 & $-0.37$ &  0.04 & $-0.40$ &  0.03 & 15.1 & ITER \\
PM J2206+4322E & 5652 &  44 &  4.62 &  0.06 & $+0.30$ &  0.03 & $+0.15$ &  0.03 & $+0.37$ &  0.03 & $+0.22$ &  0.06 & $+0.29$ &  0.02 & $+0.33$ &  0.03 &  6.2 & VESTA \\
HIP 114156 & 4314 &  66 &  4.72 &  0.06 & $-0.04$ &  0.08 & $-0.06$ &  0.05 & $-0.37$ &  0.15 & $-0.08$ &  0.05 & $-0.24$ &  0.07 & $-0.15$ &  0.03 & 15.5 & ITER \\
HIP 117960 & 5252 &  44 &  4.54 &  0.06 & $+0.12$ &  0.03 & $+0.06$ &  0.03 & $+0.14$ &  0.03 & $+0.04$ &  0.05 & $+0.12$ &  0.02 & $+0.11$ &  0.03 &  4.0 & ITER \\
HIP 118282 & 5178 &  44 &  4.64 &  0.06 & $-0.67$ &  0.03 & $-0.47$ &  0.03 & $-0.54$ &  0.03 & $-0.36$ &  0.05 & $-0.35$ &  0.03 & $-0.57$ &  0.03 &  2.0 & ITER \\
\hline
\end{tabular}
 \end{centering}
{\footnotesize $^a$ ITER: parameters determined using {\it Hipparcos} parallaxes and $Y^2$ isochrones. VESTA: parameters determined using classical SME fitting (no parallax information included) with a correction using Vesta as described in \citet{2005ApJS..159..141V}.}
\label{tab:primaries}
\end{table*}

Our analysis of the ESPaDOnS spectra is designed to keep our metallicities consistent with those from the SPOCS catalog (both are based on SME analysis and use the same set of spectral lines). As an extra check on consistency we have obtained CFHT spectra for three stars in the SPOCS sample. The derived stellar parameters from these three spectra are consistent (within errors) with those listed in the SPOCS catalog, confirming that there is no systematic offset between metallicities from SPOCS and CFHT. 

\begin{figure}
\centering
\includegraphics[width=8cm]{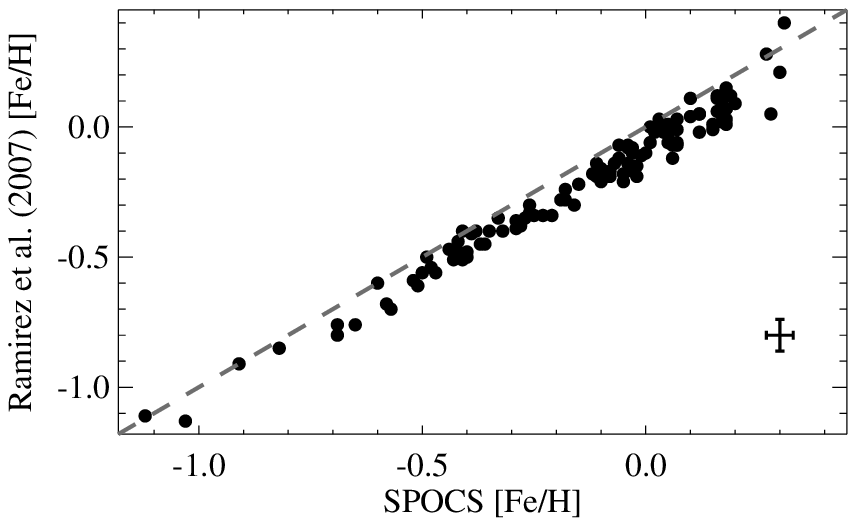}
\caption{Comparison of metallicities from \citet{2007A&A...465..271R} and those from SPOCS \citep{2005ApJS..159..141V} for the 112 stars present in both samples. The [Fe/H] values are quite consistent once corrected for a systematic offset of 0.08~dex. The remaining scatter is only $\simeq0.06$~dex. Assuming SPOCS [Fe/H] values are accurate to 0.03~dex, this implies [Fe/H] metallicities from \citet{2007A&A...465..271R} are accurate to $\simeq 0.05$~dex. We perform a similar analysis for all other metallicity sources (see Table~\ref{tab:metalsources}).}  \label{fig:control}
\end{figure}

Metallicities from other literature sources are not necessarily determined in the same way as our spectral analysis, and thus may have small systematic inconsistencies. We correct for this by checking for overlap between the SPOCS samples and any given literature source. For us to use a metallicity derived from any other literature source we require; (1) at least 30 stars in both samples that can be used as a control sample to check for differences, (2) metallicities for our primary stars from the literature source fall within the range of metallicities of the control sample, (3) the mean difference between the SPOCS metallicities and the literature metallicities in the control sample $\Delta$[Fe/H]$_{\mathrm{control}}$ $\le0.07$~dex, and (4) the resulting scatter in the control sample $\sigma_{\mathrm{control}}$ $\le0.08$~dex. These limits are designed to keep uncertainties in the primary star metallicities well below the precision already obtained for determining M dwarf metallicities (e.g., R12 and T12). We adopt $\sigma_{\mathrm{control}}$ as the uncertainty in [Fe/H] for a given literature source. As an example, we show metallicities from both \citet{2007A&A...465..271R} and SPOCS in Figure~\ref{fig:control}. We list all sources of metallicities, the adopted systematic offset (which we apply for all calculations in this paper) for that source, and the adopted uncertainty in Table~\ref{tab:metalsources}.

\begin{table} 
\setlength{\tabcolsep}{0.04in}
\caption{Corrections Applied to Primary Star Metallicities}
\centering
\begin{tabular}{l l l l l l l l l l l}
\hline \hline
Source & Control Stars & $\Delta \mathrm{[Fe/H]}$ & $\sigma_{\mathrm{control}}$ & No. of Stars Used$^a$  \\  \hline
VF05 & ... & 0.00 &  0.03 &  33\\
CFHT & ... & 0.00 &  0.03$^b$ &  50\\
C01 & 294 &  0.02 &  0.07 &   1\\
M04 & 178 &  0.04 &  0.08 &   1\\
LH05 & 174 &  0.02 &  0.06 &   1\\
T05 & 127 &  0.01 &  0.05 &   1\\
B06 &  33 &  0.07 &  0.07 &   1\\
Ra07 & 112 &  0.08 &  0.05 &   4\\
Ro07 & 127 &  0.00 &  0.07 &   5\\
F08 & 165 &  0.03 &  0.06 &   2\\
C11$^c$ & 614 &  0.00 &  0.08 &   10\\
S11 &  50 &  0.03 &  0.04 &   1\\
N12$^d$ & 125 &  0.00 &  0.05 &   2\\
\hline
\end{tabular}
\flushleft
 {\footnotesize See Table~\ref{tab:sample} for the abbreviations on references.}
 {\footnotesize $^a$Number of stars from listed source used in our final wide binary sample.}\\
 {\footnotesize $^b$Typical uncertainty. Errors for individual stars listed in Table~\ref{tab:primaries}.}\\
 {\footnotesize $^c$Metallicities determined from Str\"{o}mgren photometry.}\\
 {\footnotesize $^d$Based on a control sample from \citet{2004A&A...415.1153S} and \citet{2011A&A...533A.141S}, on which abundances in N12 are anchored.}\\
\label{tab:metalsources}
\end{table}

\subsection{Late-K/M Dwarf Spectral Types}
We determine M dwarf spectral types using indices at both visible and NIR wavelengths. We use the empirical spectral type--band index relations from \citet{Lepine:2013lr}, which have been calibrated to work on the SNIFS/UH2.2m. \citet{Lepine:2013lr} determined spectral types accurate to $\simeq 0.2$  subtypes based on empirical relations between spectral type and the strengths of TiO and CaH bands \citep{Reid:1995lr}. It has been shown that CaH is sensitive to spectral type, and that TiO is sensitive to both spectral type and metallicity \citep{Woolf:2006uq}. As a result, for stars with [Fe/H]$<-0.5$ (the metallicity of the primary star) we base our visible wavelength spectral types solely on relations using CaH bands. 

R12 showed that one can determine temperatures and spectral types using a modified version of the H$_2$O-K \citep[H$_2$O-K2,][]{2010ApJ...722..971C} index. R12 calibrated their spectral types based on $K$-band spectra of stars from the Research Consortium on Nearby Stars Measuring \citep[RECONS;][]{Henry:1994fk}. Their calibration is accurate to $\simeq0.6$ subtypes. 

\begin{figure}
\centering
\includegraphics[width=8cm]{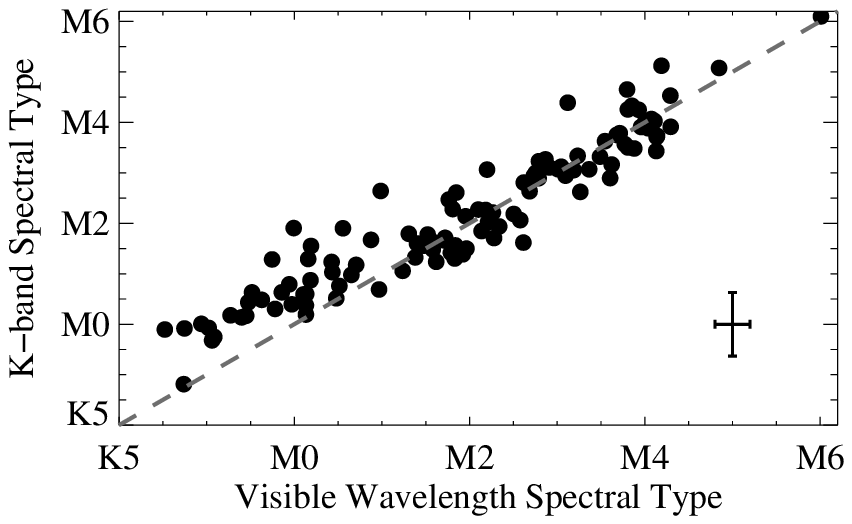}
\caption{Spectral types derived from the H$_2$O-K2 index (RA12) vs. those derived from TiO and CaH indices at visible wavelengths. The dashed line indicates a perfect agreement and typical errors are shown in the bottom right of the plot. For spectral types later than M1 the different techniques agree, but for earlier-type stars the H$_2$O-K2 index tends to return a later spectral type than visible wavelength indices.  \label{fig:sptype}}
\end{figure}

Figure~\ref{fig:sptype} compares the spectral types derived from visible wavelength indices versus those derived using the H$_2$O-K2 index. Although there is good agreement between the two techniques for the later-type stars in our sample, for stars earlier than M1 (as determined by TiO and CaH bands), spectral types determined from H$_2$O-K2 are systematically later than those from visible wavelength indices. The H$_2$O features become quite weak in the spectra of late-K and early-M stars and R12 caution using it on stars with \teff$>4000$. Further, the spectral-type calibration from R12 does not include any K stars, and is therefore unreliable for the warmest stars in our sample. As a result, we choose to use spectral types determined from our visible wavelength spectra for the entire sample.

\section{Identifying Metal-Sensitive Indices}\label{sec:MCanalysis}
\begin{figure*}
\centering
\includegraphics[width=18cm]{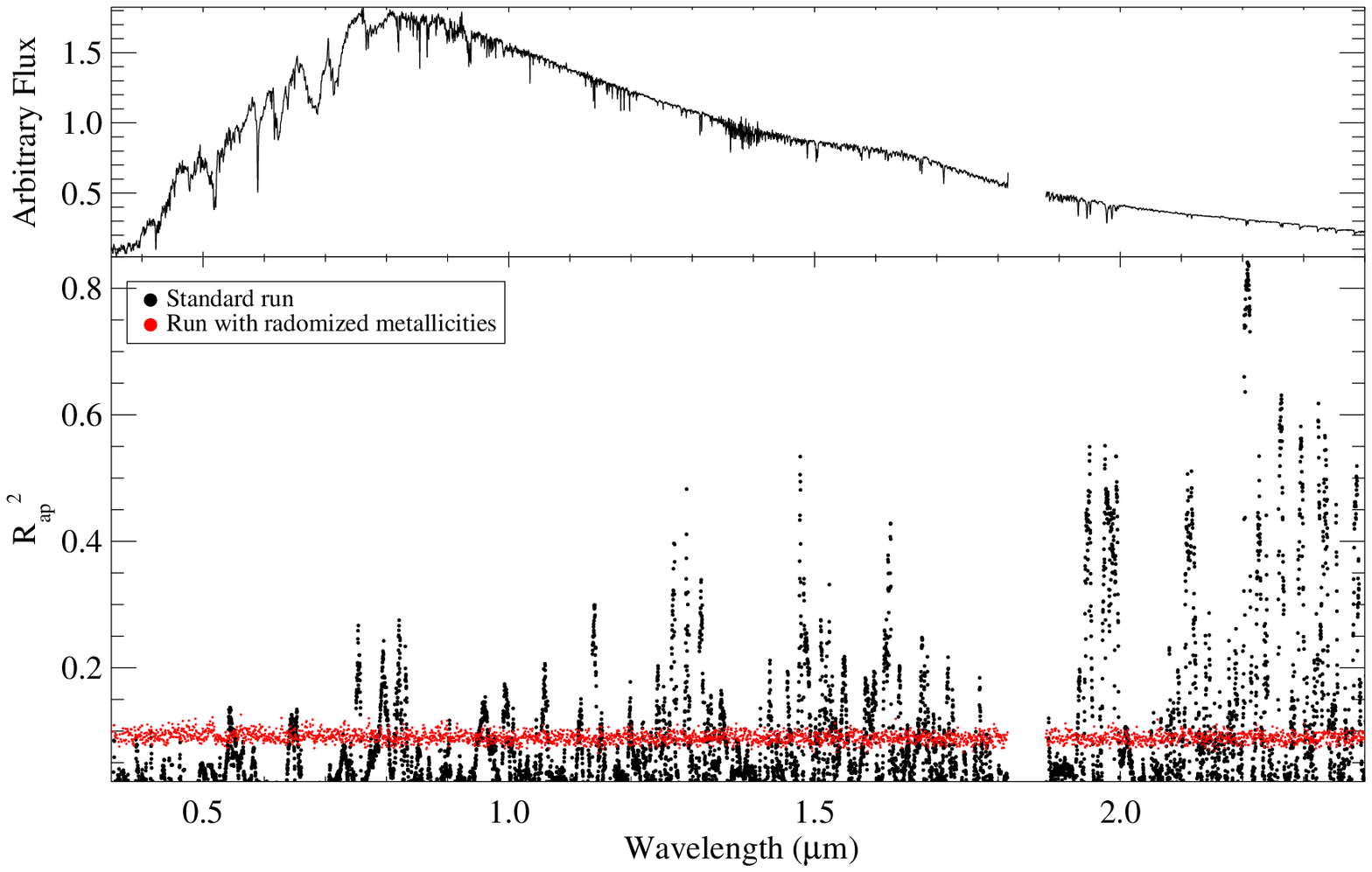}
\caption{Results of our systematic search to find the features that best correlate with [Fe/H] in late-K or M dwarf spectra. Feature centers (shown on the $X$ axis) and widths are changed incrementally covering a range of widths (20\,\AA\ to 100\,\AA) and feature centers (0.35\,\um\ to 2.4\,\um). Red points indicate the R$^2_{\mathrm{rand}}$ values derived from randomly reassigning primary star metallicities and repeating our process 1000 times (the 99.9\% highest resulting R$_{\mathrm{ap}}^2$ values are shown). In this particular analysis, we fit for [Fe/H], and fit the pseudo-continuum using just the continuum regions immediately blueward and redward of a given feature. A range of feature widths are shown, which results in some range to the distribution for a given central wavelength. Although there are $>400,000$ unique combinations of feature center and width, only 15,000 are shown for simplicity. \label{fig:MCrun1}}
\end{figure*}

To determine which features in the companion dwarf spectra best correlate with metallicity we perform a systematic analysis of our sample of spectra and metallicities. Our analysis proceeds as follows:
\begin{enumerate} 
\item A center wavelength is selected, starting at the blue end of the spectrum ($\simeq0.33$\,\um) and incrementally increasing by 0.00015\,\um\ (1.5\,\AA) after all other steps are complete. This process is repeated until the center is at the red end of the spectrum ($\simeq2.4$\,\um) and excludes the gap in all SpeX spectra at 1.85\,\um. 
\item For each feature center, we select a feature width starting at 0.002\,\um\ (20\,\AA), and then increased incrementally by 0.00015\,\um\ (1.5\,\AA) after completing all following steps. We use 20\,\AA\ as a minimum, as features smaller than this have considerable Poisson noise (making their measurement difficult). We use an upper limit of 0.01\,\um\ (100\,\AA) for the feature width, as regions of the spectra larger than this likely contain multiple features that should be treated separately. 
\item The equivalent width is calculated for each feature using the approximation:
\begin{equation}\label{eqn:eqw}
\mathrm{EW}_\lambda \simeq \sum_{i=0}^{n-1}\left[1 - \frac{F(\lambda_i)}{F_c(\lambda_i)} \right]\Delta \lambda_i,
\end{equation}
where $\lambda_i$ is the wavelength at pixel $i$, $F$ is the flux at $\lambda_i$, $F_c$ is the pseudo-continuum at $\lambda_i$, and the sum is computed over all $n$ pixels within a given feature. We compensate for the low resolution by interpolating the spectrum near the edge to a much higher resolution ($R>10,000$). We experiment with different techniques to calculate the pseudo-continuum (see below). The list of continuum regions used is listed in Table~\ref{tab:contregions}, many of which are taken from T12.
\item A temperature-sensitive parameter $\tau$, is calculated from each spectrum. $\tau$ is defined based on the center wavelength of the selected feature from step (2). Specifically, $\tau$ is set to be the $K$-band H$_2$O (H$_2$O-K2) index as defined by R12 if the feature is centered in the $K$ band or the $H$ band H$_2$O (H$_2$O-H) index as defined by T12 if the feature is in the $H$ band. If the feature falls in visible wavelengths we use a version of the Color1 index from \citet{Hawley:2002fk}, which is defined as:
\begin{equation}\label{eqn:color1}
\mathrm{Color1} = \frac{\sum_{\lambda=0.89\mu m}^{\lambda=0.9\mu m}F(\lambda)}{\sum_{\lambda=0.735\mu m}^{\lambda=0.745\mu m}F(\lambda)}.
\end{equation}
If the feature falls within the J-band, we use a new H$_2$O-J index defined as:
\begin{equation}\label{eqn:h20j}
\mathrm{H}_2\mathrm{O} \mathit{-} \mathrm{J} = \frac{\langle \mathcal{F}(1.210-1.230) \rangle/\langle \mathcal{F}(1.313-1.333) \rangle}{\langle \mathcal{F}(1.313-1.333) \rangle/\langle \mathcal{F}(1.331-1.351) \rangle},
\end{equation}
where $\langle\mathcal{F}(a-b) \rangle$ indicates the median flux level in a wavelength range between $a$ and $b$ (in \um). H$_2$O-J is defined to select regions relatively clear of atomic or molecular features and to correlate well with the H$_2$O-K2 and H$_2$O-H indices.
\item Using least-squares, the best fit is found for the equation:
\begin{equation}\label{eqn:fit}
\mathrm{[Fe/H]}_{i} = A + B\times \mathrm{EW}_{i} + C\times \tau_{i},
\end{equation}
where [Fe/H]$_i$ is the metallicity of $i$th primary star (assumed to be the metallicity of the companion M dwarf), EW$_i$ is the calculated equivalent width of the selected feature in the $i$th late-K or M dwarf companion spectrum, and $A, B,$ and $C$ are fitting parameters. The quality of the fit is measured by the adjusted square of the multiple correlation coefficient ($R_{\mathrm{ap}}^2$), which is defined as:
\begin{eqnarray}\label{eqn:rapsq}
R_{\mathrm{ap}}^2 = 1 - \frac{(n - 1)\sum(y_{i,\mathrm{model}}- {y_i})^2}{(n-p)\sum(y_i - \bar{y})^2},\\ \nonumber
\end{eqnarray}
where $p$ is the number of changeable parameters (i.e. A, B, and C), $n$ is the number of data points in the fit, $y_i$ is the metallicity of the $i$th primary star, $y_{i,\mathrm{model}}$ is the metallicity of the $i$th star predicted by the fit, and $\bar{y}$ is the average of $y_i$. A R$_{\mathrm{ap}}^2$ closer to 1 implies that the model accurately explains the variance of the sample, while R$_{\mathrm{ap}}^2$=0 implies that it can explain none. For Equation~\ref{eqn:fit}, $p=3$, $y_i$=[Fe/H]$_i$, and $y_{i,\mathrm{model}}=A + B\times \mathrm{EW}_{i} + C\times \tau_{i}$. Note that for $n\gg p$, R$_{\mathrm{ap}}^2 \simeq R^2$. 
\item To asses the significance of the assigned R$_{\mathrm{ap}}^2$ value, [Fe/H] (or [M/H]) values are randomly reassigned among the stars, and step 5 is repeated 1000 times (re-randomizing the metallicities each time). The resulting distribution of the 1000 R$_{\mathrm{ap}}^2$ values gives the level above which the R$_{\mathrm{ap}}^2$ value (determined from non-random metallicities) can be considered significant. We consider the given feature center to be a {\it bona-fide} metal sensitive feature if R$_{\mathrm{ap}}^2$ is higher than the 99.9\% highest R$_{\mathrm{ap}}^2$ value from the randomly assigned metallicities (henceforth R$_{\mathrm{rand}}^2$).
\end{enumerate}
For each increment in feature center and width, we record the resulting R$_{\mathrm{ap}}^2$ and R$_{\mathrm{rand}}^2$. We show the resulting distribution of R$^2$ and R$_{\mathrm{rand}}^2$ values as a function of the feature's central wavelength in Figure~\ref{fig:MCrun1}.

\begin{table} 
\caption{Continuum Regions Used}
\centering
\begin{tabular}{ccccccc}
\hline \hline
Visible & $J$-Band & $H$-Band & $K$-Band \\ 
 \,(\um) & \,(\um) & \,(\um) & \,(\um) \\ 
\hline
0.4035--0.4080 & 0.9790--0.9890 & 1.4440--1.4480 & 1.8860--1.8900 \\ 
0.4135--0.4180 & 1.0610--1.0650 & 1.4644--1.4710 & 1.9350--1.9400 \\ 
0.4425--0.4450 & 1.1260--1.1300 & 1.4921--1.4965 & 1.9610--1.9700 \\ 
0.4610--0.4625 & 1.1530--1.1580 & 1.5060--1.5090 & 2.0500--2.0540 \\ 
0.4680--0.4700 & 1.1890--1.1930 & 1.5190--1.5220 & 2.0800--2.0870 \\ 
0.5269--0.5299 & 1.2140--1.2180 & 1.5920--1.5960 & 2.1330--2.1351 \\ 
0.5660--0.5675 & 1.2250--1.2300 & 1.6230--1.6310 & 2.1530--2.1590 \\ 
0.6586--0.6607 & 1.2550--1.2634 & 1.6935--1.6980 & 2.1670--2.1720 \\ 
0.7041--0.7049 & 1.2700--1.2730 & 1.7530--1.7570 & 2.1940--2.1985 \\ 
0.7390--0.7500 & 1.2950--1.2970 & ... &2.2130--2.2190 \\ 
0.8100--0.8160 & 1.3040--1.3070 & ... &2.2450--2.2520 \\ 
0.8230--0.8300 & 1.3214--1.3270 & ... &2.2717--2.2781 \\ 
0.8590--0.8620 & 1.4090--1.4150 & ... &2.2850--2.2900 \\ 
0.8890--0.8920 & ... &... &2.3050--2.3105 \\ 
0.9100--0.9120 & ... &... &2.3600--2.3640 \\ 
0.9220--0.9255 & ... &... &2.3710--2.3760 \\ 
... &... &...&2.3950--2.4050 \\ 
\hline
\end{tabular}
\label{tab:contregions}
\end{table}

We repeat our analysis using various methods of fitting for the pseudo-continuum. Better estimates of the continuum should result in more accurate line measurements, and therefore higher R$_{\mathrm{ap}}^2$ values for the same features. In one experiment we tried to fit the global spectrum with a high order ($>10$) polynomial. We tested fitting each band (visible, $JHK$) with 3rd through 6th order polynomials, as well as with 3rd through 6th order Legendre polynomials. Interestingly, we found we had the best (highest R$_{\mathrm{ap}}^2$ with respect to R$_{\mathrm{rand}}^2$) results when fitting the pseudo-continuum using a linear fit (using the IDL code linfit) of the continuum regions immediately blueward and redward of the selected feature. We use this fitting procedure for all calibrations derived in Section~\ref{sec:determinemetal}.

We run additional experiments to test the influence of spectral type on determination of metallicities: we repeat our analysis on just the early-type stars in our sample (K5.5--M2.0) and again with just the late-type stars (M2.0--M6). The split roughly corresponds to our median spectral type ($\simeq$~M2). It is possible to parse our data into smaller spectral type ranges, although this will proportionately shrink each sample. Thus this would make it difficult to identify metal sensitive features that have not been previously discovered.

Lastly, we rerun our analysis using [M/H] rather than [Fe/H] for the metallicity of the primary star. R12 were able to derive better fits between $K$-band atomic lines and [M/H] than for [Fe/H], perhaps due to variations in [$\alpha$/Fe] creating discrepancies between the actual measured quantities (Na and Ca in the case of R12) and Fe abundances. However, many of the literature sources we draw from do not provide [M/H] values for the primary stars (just [Fe/H]). As a result, there are 5 fewer binaries when using [M/H] and 16 primary stars have a different source for their [M/H] values than their [Fe/H] values (see Table~\ref{tab:sample} for a full list of [Fe/H] and [M/H] sources used).

We consider a feature to be metal-sensitive if a given feature's center and width has an R$_{\mathrm{ap}}^2$ value are above its corresponding R$_{\mathrm{rand}}^2$ value. This criterion ensures that no features are identified simply by coincidence.

Once features are identified, we then attempt to derive a calibration for a given wavelength range (e.g., J-band) by solving the equation (by least squares):
\begin{equation}\label{eqn:multifit}
\mathrm{[Fe/H]}_i =  A +  \sum_{j=1}^{N}(B_j \times \mathrm{EW}_{i,j})  + C \times \tau_i ,
\end{equation}
where [Fe/H]$_i$ is the metallicity of the $i$th primary star, $N$ is the total number of features of interest among $M_\lambda$ features identified as metal-sensitive in a given wavelength range, $\tau_i$ is the temperature sensitive parameter selected based on the wavelength regime (see above), EW$_{i,j}$ is the equivalent width of the $j$th feature measured for the $i$th star, and A, C, and the $B_j$'s are fitting parameters. We find the best fit for $N$=1,2,3,...$<M_\lambda$ until the increase in R$_{\mathrm{ap}}^2$ is negligible ($\Delta R_{\mathrm{ap}}^2 <$ 0.03) or it is clear from visual inspection of the data that the adding of further variables is over fitting the data. This limit is usually hit at $N = 3--4$. Although we use [Fe/H] in Equations~\ref{eqn:fit} and \ref{eqn:multifit}, we also perform the same procedure using [M/H].

\section{Determination of M dwarf Metallicities}\label{sec:determinemetal} 
The first thing our analysis gives us is a catalog of metal-sensitive features, which we list in order of $\lambda_c$ in Table~\ref{tab:features}. In total we find 120 features that are statistically significant predictors of metallicity, although only 20 of these are used in our final calibrations. We identify a number of previously known metal-sensitive features, as well as many of new ones. One of the most metal-sensitive features is the Na~I doublet in the $K$-band (2.208\,\um), already identified by R10. Our analysis identifies the Ca~I (1.616\,\um and 1.621\,\um) and K~I (1.5176\,\um) lines shown to be metal-sensitive by T12. In fact, our analysis locks on to very similar wavelength centers and widths as those found by T12 for both $H$-band and $K$-band features. Since our analysis covers all wavelengths and is completely blind (e.g., they have no a-priori line lists or knowledge of feature size) this suggests that our purely empirical analysis is identifying metal-sensitive atomic and molecular lines. 

\begin{table} 
\begin{centering}
\caption{Metal-sensitive Features}
\renewcommand{\arraystretch}{0.4}
 \setlength{\tabcolsep}{0.04in}
\begin{tabular}{lll | ccc | ccc}
\hline \hline
\multicolumn{1}{c}{F\#} & \multicolumn{1}{c}{Center $\lambda$} & \multicolumn{1}{c|}{Width}   & \multicolumn{6}{c}{R$_{\mathrm{ap}}^2$} \\
 & \multicolumn{1}{c}{ [$\mu$m]} & \multicolumn{1}{c|}{ [\AA]} & \multicolumn{3}{c}{[Fe/H]} & \multicolumn{3}{c}{[M/H]} \\
&  &  & \colhead{All} &  \colhead{Early} &  \colhead{Late}  & \colhead{All} &  \colhead{Early} &  \colhead{Late}  \\
 \hline
F01 & 0.4648 &  23 & ... & 0.56 & ... & ... & 0.54 & ... \\
F02 & 0.5608 &  20 & ... & 0.43 & ... & ... & 0.30 & ... \\
F03 & 0.6118 &  20 & 0.14 & 0.20 & 0.26 & 0.15 & 0.26 & 0.39 \\
F04 & 0.6232 &  20 & ... & ... & 0.28 & ... & ... & 0.41 \\
F05 & 0.6416 &  41 & 0.13 & ... & 0.42 & 0.22 & ... & 0.57 \\
F06 & 0.7540 &  20 & 0.30 & 0.51 & 0.39 & 0.28 & 0.44 & 0.38 \\
F07 & 0.8208 &  35 & 0.36 & 0.77 & 0.23 & 0.31 & 0.69 & 0.22 \\
F08 & 0.8684 &  26 & 0.14 & 0.63 & 0.23 & 0.15 & 0.53 & 0.18 \\
F09 & 1.1396 &  26 & 0.35 & 0.46 & 0.28 & 0.27 & 0.35 & 0.19 \\
F10 & 1.2698 &  98 & 0.42 & 0.47 & 0.37 & 0.43 & 0.47 & 0.35 \\
F11 & 1.2908 &  20 & 0.51 & 0.63 & 0.41 & 0.41 & 0.55 & 0.26 \\
F12 & 1.3148 &  50 & 0.39 & 0.72 & 0.23 & 0.31 & 0.66 & 0.25 \\
F13 & 1.3344 &  23 & 0.15 & 0.42 & 0.31 & 0.16 & 0.35 & 0.38 \\
F14 & 1.4766 &  41 & 0.54 & 0.60 & 0.44 & 0.45 & 0.51 & 0.29 \\
F15 & 1.4836 &  23 & 0.43 & 0.73 & 0.18 & 0.36 & 0.66 & 0.25 \\
F16 & 1.5172 &  33 & 0.48 & 0.70 & 0.48 & 0.48 & 0.69 & 0.45 \\
F17 & 1.6158 &  23 & 0.60 & 0.88 & 0.28 & 0.55 & 0.86 & 0.22 \\
F18 & 1.7261 &  32 & 0.24 & 0.39 & 0.26 & 0.18 & 0.38 & 0.19 \\
F19 & 2.2079 &  68 & 0.86 & 0.88 & 0.83 & 0.78 & 0.72 & 0.68 \\
F20 & 2.3242 &  38 & 0.63 & 0.75 & 0.58 & 0.56 & 0.57 & 0.50 \\
F21 & 2.3342 &  35 & 0.61 & 0.73 & 0.55 & 0.57 & 0.58 & 0.49 \\
F22 & 2.3844 &  35 & 0.54 & 0.69 & 0.35 & 0.50 & 0.51 & 0.32 \\
\hline
\end{tabular}
\label{tab:features}
\end{centering}
 {\footnotesize Here we show only the features used in our final calibrations. The full version of the table with {\it all} metal-sensitive features identified by our analysis will be available electronically.}\\
 {\footnotesize $^\mathrm{a}$... denotes that the feature did not have an R$_{\mathrm{ap}}^2$ value above the R$_{\mathrm{rand}}^2$ value, and thus is not considered a statistically significant metal-sensitive feature. }\\
 {\footnotesize $^\mathrm{b}$Full: K5.5--M6.0, early: K5.5--M2.0, late: M2.0--M6.0.}
\end{table}

Because our only restriction was the size of the feature ($\le 100$\,\AA), our technique can easily identify areas of the spectrum corresponding to several, even unrelated lines. Some of these regions may be associated with doublets/triplets from the same atomic species, with broad molecular bands, or with sets of lines that are blended at our resolution. Some features in Table~\ref{tab:features} may not correspond to any one specific element or molecule, but simply to a region of the spectrum that undergoes overall changes as a function of the metallicity of the star.

\begin{figure*}
\centering
\includegraphics[width=8cm]{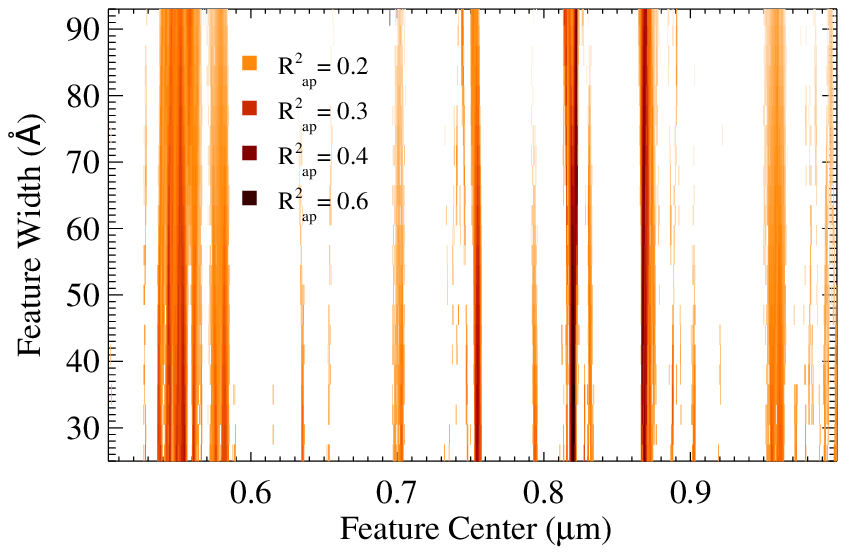}
\includegraphics[width=8cm]{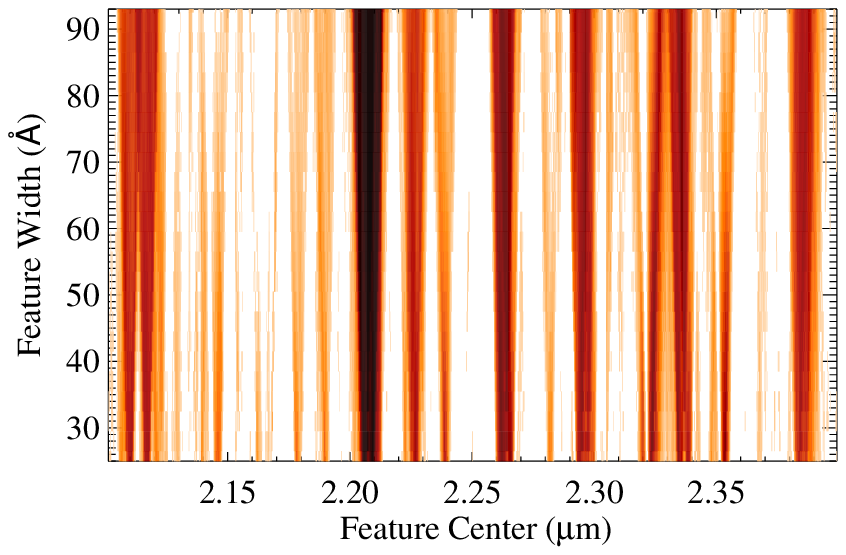}
\caption{Distribution of R$_{\mathrm{ap}}^2$ values as a function of feature center and width for the $K$-band (left) and red end of the visible wavelengths (right). The visible wavelength distribution is based on our analysis using just K5.5--M2 dwarfs. Data points which do not have R$_{\mathrm{ap}}^2>$R$_{\mathrm{rand}}^2$ are white (i.e. not shown). Surprisingly, there is not much change in R$_{\mathrm{ap}}^2$ as a function of the selected width for a given feature in the $K$-band, however we do see a trend towards smaller widths in the visible wavelengths. \label{fig:width_center}}
\end{figure*}

We show the distribution of R$_{\mathrm{ap}}^2$ values as a function of feature width and center for two example wavelength regions in Figure~\ref{fig:width_center}. Interestingly, features in the $H$ and $K$ band features yield similar R$_{\mathrm{ap}}^2$ for a wide range of feature widths. However, the opposite is seen in the visible end of the spectrum, where features perform better near the minimum feature width (20\,\AA). This is most likely due to crowding at visible wavelengths. 

We find far better metal-sensitive features by doing a separate analysis for earlier-type dwarfs (K5.5--M2) and for later type dwarfs (M2-M6). Specifically, the best fit we achieve when fitting for [Fe/H] using all stars in our sample yields R$_{\mathrm{ap}}^2$=0.54, whereas when we split up the sample by spectral type we achieve R$_{\mathrm{ap}}^2$=0.84 for K5.5--M2.0 and R$_{\mathrm{ap}}^2$=0.68 for M2.0--M6.0. This is not unexpected, many of the most metal-sensitive features for K5.5-M2 dwarfs become blended with molecular bands (which grow as a function of spectral type) at the resolution of SNIFS. Further, features blueward of 0.5\, \um\ tend to have very low S/N for stars later than M2, and are not to be very useful. This also suggests that better results could be achieved on later type M dwarfs with modest improvements in resolution, to better distinguish the lines.
 
We find the best empirical fits to Equation~\ref{eqn:multifit} for each wavelength regime (visible, $J$, $H$, and $K$) and metallically metric ([Fe/H] and [M/H]). They are:
\begin{eqnarray}
\label{eqn:OP37} \mathrm{[Fe/H]_{V,e}} &=& 0.53F_{07} + 0.26F_{01} - 0.16F_{02} \\ \nonumber && - 0.784\mathrm{(Color1)} - 0.34 \\ 
\label{eqn:OP38} \mathrm{[M/H]_{V,e}} &=& 0.38F_{07} + 0.21F_{01} + 0.29F_{08} \\ \nonumber && - 0.504\mathrm{(Color1)} - 0.79 \\ 
\hline
\, \nonumber \\
\label{eqn:OP39} \mathrm{[Fe/H]_{V,l}} &=&  -0.20F_{05} + 0.48F_{08} + 0.24F_{07} \\ \nonumber && + 0.14F_{03} - 0.204\mathrm{(Color1)} - 0.32 \\ 
\label{eqn:OP40} \mathrm{[M/H]_{V,l}} &=&  -0.065F_{05} - 0.071F_{04} - 0.30F_{06} \\ \nonumber && + 0.719\mathrm{(Color1)} - 0.24 \\ 
\hline
\, \nonumber \\
\label{eqn:J33} \mathrm{[Fe/H]_J} &=& 0.29F_{10} + 0.21F_{09} + 0.26F_{12} \\ \nonumber && - 0.26F_{13} - 0.190\mathrm{(\mathrm{H}_2\mathrm{O} \mathit{-} \mathrm{J})} - 1.03 \\ 
\label{eqn:J34} \mathrm{[M/H]_J} &=& 0.32F_{10} + 0.46F_{11} + 0.076F_{09} \\ \nonumber && + 1.213\mathrm{(\mathrm{H}_2\mathrm{O} \mathit{-} \mathrm{J})} - 1.97 \\ 
\hline
\, \nonumber \\
\label{eqn:H33} \mathrm{[Fe/H]_H} &=& 0.40F_{17} + 0.51F_{14} - 0.28F_{18} \\ \nonumber && - 1.460\mathrm{(\mathrm{H}_2\mathrm{O} \mathit{-} \mathrm{H})} + 0.71 \\ 
\label{eqn:H34} \mathrm{[M/H]_H} &=& 0.38F_{17} + 0.40F_{16} + 0.41F_{15} \\ \nonumber && + 0.194\mathrm{(\mathrm{H}_2\mathrm{O} \mathit{-} \mathrm{H})} - 0.76 \\ 
\hline
\, \nonumber \\
\label{eqn:K33} \mathrm{[Fe/H]_K} &=& 0.19F_{19} + 0.069F_{22} + 0.083F_{20} \\ \nonumber && + 0.218\mathrm{(\mathrm{H}_2\mathrm{O} \mathit{-} \mathrm{K2})} - 1.55 \\ 
\label{eqn:K34} \mathrm{[M/H]_K} &=& 0.12F_{19} + 0.086F_{22} + 0.13F_{21} \\ \nonumber && + 0.245\mathrm{(\mathrm{H}_2\mathrm{O} \mathit{-} \mathrm{K2})} - 1.18 
\end{eqnarray}
where $F{\#}$ refer to the equivalent width of the corresponding feature listed in Table~\ref{tab:features}, the subscripts refer to the wavelength bands where the calibration is useful (V referring to visible wavelengths). An additional subscript is added (e or l) for calibrations in visible wavelengths to denote which formula is valid for early (K5.5 to M2) and late (M2 to M6) dwarfs. All equations assume feature equivalent widths are calculated in Angstroms. 

\begin{figure*}
\centering
\includegraphics[width=\textwidth]{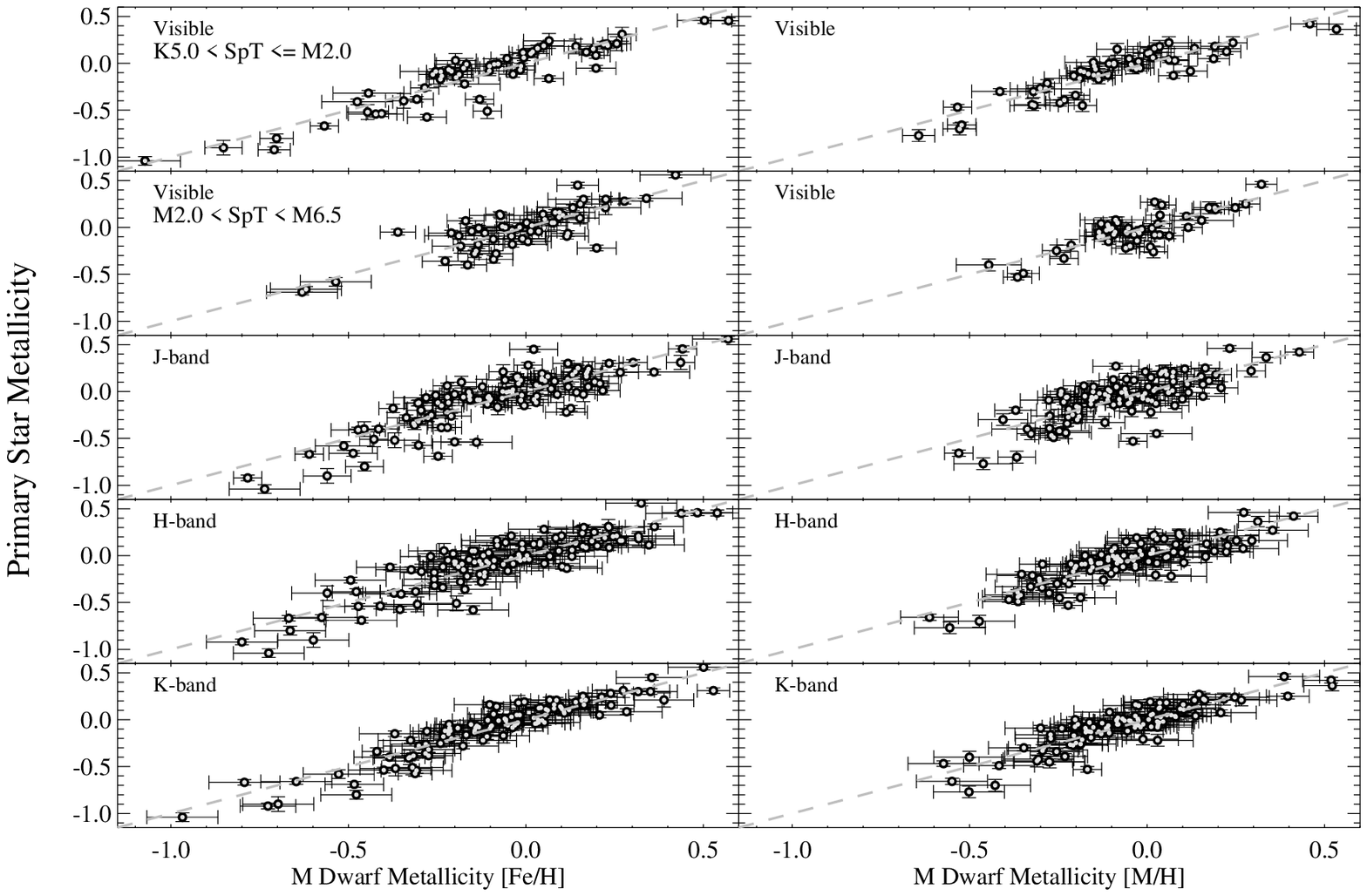}
\caption{Metallicity for the primary star vs. derived metallicity of the late K or M dwarf companion based on Equations~(\ref{eqn:OP37})-(\ref{eqn:K34}). Calibrations for [Fe/H] are shown on the left plots, while those for [M/H] are shown on the right plots. Statistics on the quality of the fit can be found in Table~\ref{tab:calibrations}. $Y$ axis error bars shown are based on 1$\sigma$ Gaussian errors for the primary star metallicity (see Section~\ref{sec:FGK}). Error bars for the K/M dwarf metallicity are the 1$\sigma$ standard deviation of 1000 recalculations of the K/M dwarf metallicity after adding noise to each spectrum consistent with its S/N.}  \label{fig:calibrations}
\end{figure*}

We show the primary star metallicity as a function of the derived metallicity for the companion dwarf for each of the 10 calibrations in Figure~\ref{fig:calibrations} and list reduced $\chi^2$, R$_{\mathrm{ap}}^2$, root mean square error (RMSE), in Table~\ref{tab:calibrations}. The RMSE indicates how useful a model is at prediction (lower numbers indicate the fit is a better predictor) and is defined as:
\begin{equation}
\mathrm{RMSE} = \sqrt{\sum_{i=0}^n\frac{(y_{i,\mathrm{model}}- {y_i})^2}{(n-p)}}.
\end{equation}
Lower R$_{\mathrm{ap}}^2$ and higher RMSE values may in part be due to differences in S/N as a function of wavelength. We estimate measurement noise sources by adding synthetic noise to each spectrum consistent with the observed S/N of that spectrum, then recalculating the metallicity of the M dwarf using the appropriate equation above. The standard deviation in the metallicity estimate from 1000 different additions of noise pattern is assumed to be the measurement error. This error is what is what we use for our calculation of the reduced $\chi^2$ for each fit. Thus the reduced $\chi^2$ values probe how much of the noise comes from measurement (errors)A reduced $\chi^2$ close to 1 would suggest that most or all of the error from the fit is due to measurement (e.g. Poisson) noise. 

\begin{table*} 
\caption{Metallicity Calibration Statistics}
\centering
\begin{tabular}{lllllllll}
\hline \hline
Equation \# &Band & Wavelength Range & SpT Range & Metal Type &R$_{\mathrm{ap}}^2$ & RMSE & $\sigma$ & $\chi_{\mathrm{Red}}^2$ \\
\hline
\ref{eqn:OP37} & Optical &   0.35$< \lambda \le$  1.00 & K5.0--M2.0 &  [Fe/H] &   0.84 &   0.10 &   0.13 &     8.8\\
\ref{eqn:OP38} & Optical &   0.35$< \lambda \le$  1.00 & K5.0--M2.0 &  [M/H] &   0.80 &   0.10 &   0.11 &     6.4\\
\ref{eqn:OP39} & Optical &   0.35$< \lambda \le$  1.00 & M2.0--M6 &  [Fe/H] &   0.68 &   0.12 &   0.14 &     7.1\\
\ref{eqn:OP40} & Optical &   0.35$< \lambda \le$  1.00 & M2.0--M6 &  [M/H] &   0.65 &   0.12 &   0.11 &     7.8\\
\ref{eqn:J33} & J &   1.00$< \lambda \le$  1.44 & K5.0--M5.0 &  [Fe/H] &   0.71 &   0.13 &   0.16 &    11.4\\
\ref{eqn:J34} & J &   1.00$< \lambda \le$  1.44 & K5.0--M5.0 &  [M/H] &   0.55 &   0.12 &   0.15 &     9.4\\
\ref{eqn:H33} & H &   1.44$< \lambda \le$  1.80 & K5.0--M5.0 &  [Fe/H] &   0.77 &   0.11 &   0.14 &     3.7\\
\ref{eqn:H34} & H &   1.44$< \lambda \le$  1.80 & K5.0--M5.0 &  [M/H] &   0.73 &   0.09 &   0.12 &     4.2\\
\ref{eqn:K33} & K &   1.80$< \lambda \le$  2.45 & K5.0--M5.0 &  [Fe/H] &   0.86 &   0.08 &   0.11 &     4.7\\
\ref{eqn:K34} & K &   1.80$< \lambda \le$  2.45 & K5.0--M5.0 &  [M/H] &   0.77 &   0.08 &   0.10 &     3.8\\
\hline
\end{tabular}
\label{tab:calibrations}
\end{table*}

The dependence on $\tau$ varies significantly between equations. This is most likely due to different features capturing some of the temperature-dependence and/or that the Color1 and H$_2$O indices are not accurately modeling temperature dependencies across the full sample. It is also interesting that some coefficients of features are negative. This may be due to a combination of factors, including changes in [Fe/H] versus [$\alpha$/Fe] (e.g. Ca is an $\alpha$ element) or complex relations between \teff and [Fe/H], that vary for each feature. Whatever the case, since these fits are purely empirical, we should be cautious not to over-interpret the physical meaning of any particular coefficient or feature.

\section{Assessing and Recalibrating Existing Techniques}\label{sec:recal}
In addition to defining our own metallicity calibrations, we can use our sample to test existing metallicity estimators, as well as improve the existing calibrations. Like before, we use R$_{\mathrm{ap}}^2$, andRMSE as our standard metrics to asses the quality of a calibration. We summarize our refits in Table~\ref{tab:recalibrations}.

\begin{table} 
\caption{Assessment of Previous Metallicity Indicators}
\centering
\begin{tabular}{lllllllllllll}
\hline \hline
Technique & Type & R$_{\mathrm{ap}}^2$ & RMSE & Eqn No. \\
\hline
$\zeta_{\mathrm{TiO/CaH}}$  & [Fe/H] & 0.58 & 0.28 & \ref{eqn:zeta3feh} \\
$\zeta_{\mathrm{TiO/CaH}}$  & [M/H] & 0.61 & 0.23 & \ref{eqn:zeta3mh} \\
$J-K$ & [Fe/H] & 0.30 & 0.19 &  \ref{eqn:jkfit1} \\
$J-K$ & [M/H] & 0.25 & 0.16 &  \ref{eqn:jkfit2} \\
$H$-Band & [Fe/H] & 0.74 & 0.14 & \ref{eqn:fehkbandtest}  \\
$H$-Band & [M/H] & 0.71 & 0.12 & \ref{eqn:mhkbandtest}  \\
$K$-Band & [Fe/H] & 0.76 & 0.14 &  \ref{eqn:hfhfit} \\
$K$-Band & [M/H] & 0.75 & 0.13 & \ref{eqn:hmhfit} \\
\hline
\end{tabular}
\label{tab:recalibrations}
\end{table}

\subsection{$\zeta_{\mathrm{TiO/CaH}}$} 
Much effort has gone into determine M dwarf metallicities using visible wavelength spectra. Most of this has been focused on the $\zeta_{\mathrm{TiO/CaH}}$ (henceforth $\zeta$) parameter \citep[e.g.][]{Lepine:2007fk, Woolf:2009qy, Dhital:2012lr}. However, the setup of our analysis means that we would not be able to identify $\zeta$ at all, because $\zeta$ is based on spectroscopic indices (not equivalent widths). Band indices (e.g., TiO5, CaH3, etc.) are calculated from the ratio of the flux in region $a$ to the flux in region $b$ using the approximation:
\begin{equation}\label{eqn:index}
\mathrm{Index} \simeq \frac{[\sum_{i=a}F(\lambda_i)]/[w_a]}{[\sum_{i=b}F(\lambda_i)]/[w_b]},
\end{equation}
where $w_a$ and $w_b$ are the widths of region $a$ and $b$ in angstroms, respectively. The sums are computed over all pixels $i$ in region $a$ and $b$, respectively. Our analysis only makes use of equivalent widths (see Equation~\ref{eqn:eqw}). Further, we do not allow high order terms, while $\zeta$ generally requires 3rd or 4th order polynomials of the CaH index \citep[e.g.][]{Lepine:2013lr}. However, this does not prevent us from using our data to test the performance of $\zeta$. 

\begin{figure}
\centering
\includegraphics[width=8cm]{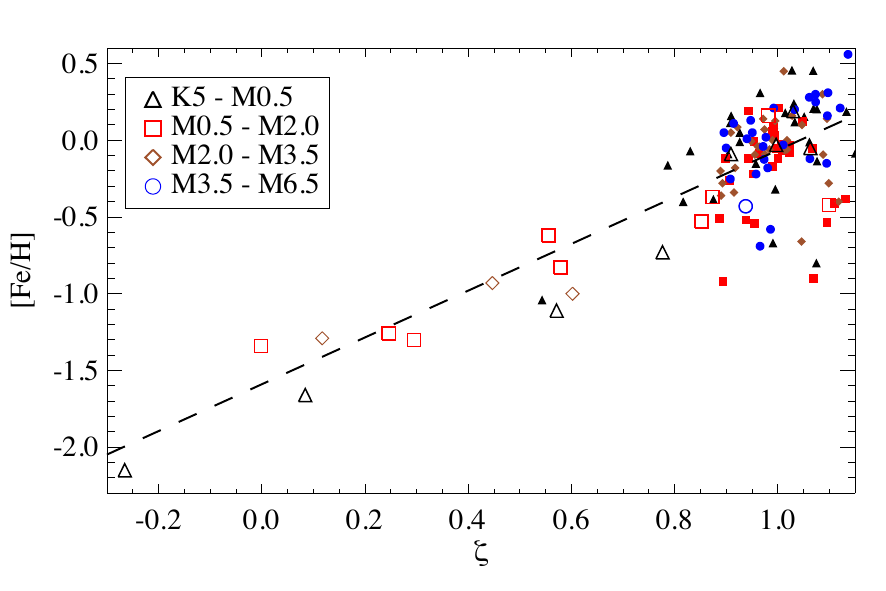}
\includegraphics[width=8cm]{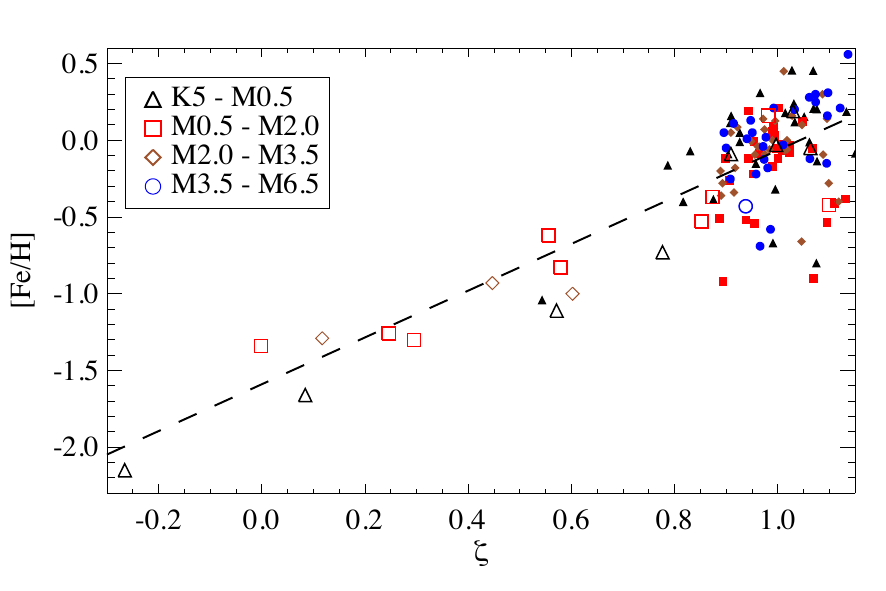}
\caption{ Metallicity of the primary star ([Fe/H] on top and [M/H] on bottom) vs. the $\zeta$ parameter as defined by \citet{Lepine:2007fk} and using the calibration of \citet{Lepine:2013lr}. It has been shown that the effectiveness of $\zeta$ varies as a function of spectral type, so we break up our sample into four spectral type ranges in the plot. The best-fit lines from Equations~\ref{eqn:zeta2feh} and \ref{eqn:zeta2mh} are shown as dotted lines. We add in a sample of 22 stars from \citet[][open symbols]{Woolf:2009qy} to our own wide binary sample (filled points). Our measurements of $\zeta$ do accurately identify metal-poor stars identified in \citet{Woolf:2009qy}, however, $\zeta$ only identifies one of our wide binary stars as a subdwarf.  \label{fig:zetas}} 
\end{figure}

We calculate the CaH2, CaH3, and TiO5 indices following the definitions from \citet{Reid:1995lr}. We compute corrected indices (CaH2$_{c}$, CaH3$_{c}$, and TiO5$_{c}$) using the formula from \citet{Lepine:2013lr}, which include corrections for the SNIFS instrument. We use these to compute $\zeta$ following the formula as defined by \citet{Lepine:2007fk}:
\begin{equation}
\zeta = \frac{1-\mathrm{TiO5}}{1-\mathrm{[TiO5]}_{Z_\odot}},
\end{equation}
where $[TiO5]_{Z_\odot}$ is a function of CaH = CaH2 + CaH3. We use the formula for [TiO5]$_{Z_\odot}$ from \citet{Lepine:2013lr}.
\begin{eqnarray}\nonumber \label{eqn:TiO5z}
\mathrm{[TiO5]}_{Z_\odot} &=& 0.622 - 1.906(\mathrm{CaH_{c}}) \\
	&&+ 2.211(\mathrm{CaH_{c}})^2 - 0.588 (\mathrm{CaH_{c}})^3.
\end{eqnarray}
We plot the primary star metallicities as a function of the derived $\zeta$ values as filled points in Figure~\ref{fig:zetas}. $\zeta$ shows a weak trend with metallicity in both [Fe/H] and [M/H].  From this we derive the following relationships:
\begin{eqnarray}
\mathrm{[Fe/H]} &=& 0.98\zeta - 1.04 \\  \label{eqn:zeta1feh}
\mathrm{[M/H]} &=&  0.68\zeta - 0.74  \label{eqn:zeta1mh} 
\end{eqnarray}
Like before, we randomly reassign the metallicities to different CPM pairs, and attempt to compute an R$_{\mathrm{rand}}^2$ value. We find that both Equations~\ref{eqn:zeta1feh} and \ref{eqn:zeta1mh} give R$_{\mathrm{ap}}^2 < $R$_{\mathrm{rand}}^2$. Further, the $\zeta$ parameter only correctly identifies one companion as a subdwarf (LHS 1812/PM I06032+1921S). Although this is the most metal-poor star in our sample, there are 12 other stars in our sample with [Fe/H]$<-0.5$ but $\zeta$ values consistent with solar metallicity ($\zeta>0.825$). According to \citet{Woolf:2009qy}, M dwarfs with [Fe/H]$<-0.34$ should have $\zeta<0.825$, and be labeled as sdM, suggesting a problem with these stars. If we remove these pairs, we derive the following relations:
\begin{eqnarray}
\mathrm{[Fe/H]} &=& 1.26\zeta - 1.25 \\  \label{eqn:zeta2feh}
\mathrm{[M/H]} &=&  0.88\zeta - 0.89  \label{eqn:zeta2mh} 
\end{eqnarray}
These formulae are highly significant; they yield R$_{\mathrm{ap}}^2$ values of 0.58 and 0.52, and RMSE values of 0.22 and 0.20, respectively. This suggests that $\zeta$ may be useful at predicting metallicities for [Fe/H]$>+0.05$ (the limit of the \citet{Woolf:2009qy} calibration), provided the high-zeta, low-metallicity stars can be explained.

\citet[][using $R\simeq3000$ spectra]{Woolf:2009qy} derive a relation between metallicity and the $\zeta$ parameter using a mix of wide binaries (for which the primary star metallicity is known) and single (K/M) stars with high-resolution spectra, analyzed using the MOOG software \citep{1973PhDT.......180S} with NEXTGEN models \citep{1999ApJ...512..377H}. Although many of their wide binaries are also in our binary sample (including LHS 1812), there is insufficient overlap between the stars in \citet{Woolf:2005fj, Woolf:2006uq}, and \citet{Woolf:2009qy} and those from SPOCS or our CFHT samples to detect any systematic offsets between the two samples. This is further complicated by fact that \citet{Woolf:2009qy} have very few subdwarf binaries in their sample (most of their subdwarfs are single stars). To test weather these low-metallicity, high-$\zeta$ dwarfs are anomalous, we observed an additional set of stars from \citet{Woolf:2009qy}. 

In total we observed 22 stars used in the \citet{Woolf:2009qy} calibration with SNIFS. We specifically select stars to cover a wide range of metallicities to get a wide range of $\zeta$ values. We list the 22 stars in Table~\ref{tab:woolfsample} and show them in Figure~\ref{fig:zetas} as unfilled points. These 22 points form a clear metallicity sequence, showing that the revised $\zeta$ can reproduce the results of \citet{Woolf:2009qy} and that $\zeta$ can be measured using modest resolution spectra.

\begin{table} 
\caption{Stars Observed from \citet{Woolf:2009qy}}
\begin{centering}
\begin{tabular}{lllllllllllll}
\hline \hline
Name & [Fe/H]$^a$ & $\sigma_{\mathrm{[Fe/H]}}^a$ & [M/H]$^a$ & $\zeta^b$ \\
\hline
LHS 38  & -0.43 &  0.05 & -0.40 &  0.94\\
HIP 1386  &  0.16 &  0.10 &  0.15 &  0.98\\
HIP 59514  & -0.05 &  0.08 & -0.03 &  1.06\\
HIP 89490  & -0.53 &  0.08 & -0.44 &  0.85\\
HIP 98906  & -0.62 &  0.10 & -0.52 &  0.56\\
HIP 105932  & -0.37 &  0.05 & -0.30 &  0.87\\
HD 18143B  &  0.19 &  0.11 &  0.18 &  1.03\\
HD 88230  & -0.03 &  0.18 & -0.05 &  1.00\\
HD 95735  & -0.42 &  0.07 & -0.40 &  1.10\\
GJ 129  & -1.66 &  0.05 & -1.33 &  0.08\\
GJ 1177B  & -0.09 &  0.10 & -0.06 &  0.91\\
GJ 3212  & -0.08 &  0.05 & -0.06 &  0.70\\
LHS 364 & -1.41 &  0.04 & -1.15 & 0.00\\
GJ 9722  & -0.83 &  0.04 & -0.70 &  0.58\\
LHS 174  & -1.11 &  0.05 & -0.95 &  0.57\\
LHS 182  & -2.15 &  0.03 & -1.88 & -0.27\\
LHS 491  & -0.93 &  0.08 & -0.78 &  0.45\\
LHS 3084  & -0.73 &  0.05 & -0.64 &  0.78\\
LHS 156 & -1.00 &  0.04 & -0.85 &  0.60\\
LHS 161 & -1.30 &  0.04 & -1.06 &  0.29\\
LHS 318 & -1.26 &  0.05 & -1.03 &  0.25\\
LSPMJ2205+5353 & -1.29 &  0.08 & -1.06 &  0.12\\
\hline
\end{tabular}
\end{centering}
 {\footnotesize $^a$Determined from \citet{Woolf:2005fj}, \citet{Woolf:2006uq}, or \citet{Woolf:2009qy}.}\\
 {\footnotesize $^b$Determined from SNIFS spectra.}
\label{tab:woolfsample}
\end{table}

Assuming the metallicities from \citet{Woolf:2009qy} are reliable and consistent with our own, we use the combined set of our binaries and the 22 additional late-type dwarfs from \citet{Woolf:2009qy} to derive the following relations:
\begin{eqnarray}
\mathrm{[Fe/H]} &=& 1.55\zeta - 1.62  \label{eqn:zeta3feh} \\ 
\mathrm{[M/H]} &=&  1.29\zeta - 1.35   \label{eqn:zeta3mh}
\end{eqnarray}
which we show in Figure~\ref{fig:zetas} as dashed lines. The resulting fits yield R$_{\mathrm{ap}}^2$ values of 0.58 and 0.61, respectively, both well above the R$_{\mathrm{rand}}^2$ (0.17 and 0.14, respectively). The RMSE values are 0.28 and 0.23, although it is notably higher for dwarfs with $\zeta>0.825$ and lower for dwarfs with $\zeta<0.825$. If we remove the 11 stars with [Fe/H]$<-0.5$ but $\zeta>0.825$, $\zeta$ follows a clear trend over the full range of metallicities covered. Interestingly, these 11 dwarfs appear to follow a completely different sequence in [Fe/H] (or [M/H]) versus $\zeta$, and are well separated from their single-star, metal-poor counterparts, suggesting that they are unique in some way. However, further inspection of these 11 pairs does not reveal anything that could explain their discrepancy: none exhibit H-$\alpha$ significant emission (likely inactive), they cover a wide range of spectral types (K7--M5), and they have metallicities from 5 different sources (including from SPOCS and our own CFHT spectra). We revisit the issue of these stars in Section~\ref{sec:discussion}.


\subsection{$J-K$ Metallicities}
\citet{Johnson:2012fk} find a relation between the $J - K$ and $V - K$ colors and the metallicity of M dwarfs, based in part on relations noted by \citet{Leggett:1992lr} and \citet{2005AJ....129.1483L}. They find a best fit relation of:
\begin{equation}\label{eqn:jjfit}
\mathrm{[Fe/H]} = -0.050 + 3.520\Delta(J-K),
\end{equation}
where $\Delta(J-K)$ is defined as:
\begin{displaymath}
   \Delta(J-K) = \left\{
    \begin{array}{lr}
       (J-K) - 0.835 & : V-K < 5.5\\
       (J-K) - \sum_{i=0} a_i (V-K)^i & : V-K \ge 5.5
     \end{array}
   \right.
\end{displaymath}
and $\{a\} = \{1.637, -0.2910, 0.02557\}$. \citet{Johnson:2012fk} note that this metallicity relation is only valid for stars with $-0.1 < \Delta(J - K) < 0.1$ and $V-K>3.8$, but that this technique yields metallicities accurate to $\pm0.15$~dex. When we apply these two restrictions to our sample, we have 118 M dwarfs with known metallicities for their primary stars. This includes stars without SNIFS/IRTF spectra that were therefore not included earlier analyses. \
We find a higher RMSE of 0.20~dex, and an R$_{\mathrm{ap}}^2$ of 0.20. One possible issue is the quality of $V$ magnitudes in our sample, which come from a variety of sources. However, when we remove stars with $V-K \ge 5.5$ unless they have more reliable $V$ magnitudes from {\it Tycho} \citep{2000A&A...355L..27H}, the quality of the fit does not change in any significant way (for stars with $V-K<5.5$, $\Delta(J-K)$ is independent of $V$ so these are not removed).

We attempt to improve on the calibration and derive a relation for [M/H] and find:
\begin{eqnarray}
\mathrm{[Fe/H]} &=& -0.11 + 3.14\Delta(J-K) \label{eqn:jkfit1} \\ 
\mathrm{[M/H]} &=&  -0.09 + 2.14\Delta(J-K)  \label{eqn:jkfit2} 
\end{eqnarray}
which results in a slightly improved RMSE $= 0.19$ and $0.16$, and improved R$_{\mathrm{ap}}^2=0.30$ and $0.25$ for [Fe/H] and [M/H] respectively. These R$_{\mathrm{ap}}^2$ values are significantly larger than the R$_{\mathrm{rand}}^2$ values (0.08 and 0.09). The major difference between our fit and that of \citet{Johnson:2012fk} is that they fix the constant term to -0.05 in order to keep [Fe/H] = -0.05 at $\Delta(J-K)=0$, consistent with a volume limited sample of stars from \citet{Johnson:2009fk}, whereas we make no such restrictions.


\subsection{$K$-band Metallicities}\label{sec:kband}
R10 have shown that one could derive M dwarf metallicities from $K$-band spectra using the Na~I and Ca~I lines (at 2.21\micron~and 2.26\micron). T12 refine the calibration of R12 using SpeX data, and find that metallicities derived this way are accurate to $\pm0.12$~dex. However, both R12 and T12 use relatively few wide binary pairs (18 and 22, respectively), and there is significant overlap in their two samples. Our sample has  overlap with theirs, but is large enough to serve as a robust check on their calibrations. Because the work of T12 was optimized for SpeX, we perform our test on their calibration. We follow their method as closely as possible (including altering our continuum fitting procedure to match theirs). 

We find that following the calibration of T12 yields RMSE $= 0.16$ and R$_{\mathrm{ap}}^2 = 0.69$.  We improve this calibration, and find a best fit of the form:
\begin{eqnarray}\nonumber \label{eqn:fehkbandtest}
\mathrm{[Fe/H]} &=& 0.19\times \mathrm{EW}_{Na} + 0.074 \times \mathrm{EW}_{Ca} \\
&& + 2.13\times (\mathrm{H}_2\mathrm{O} \mathit{-} \mathrm{K}) - 3.18
\end{eqnarray}
This new form yields metallicities accurate to RMSE $= 0.14$ and R$_{\mathrm{ap}}^2 = 0.76$. 
The new calibration noticeably improves the fit for stars with $-0.3 < $[Fe/H]$ < +0.0$, however, both calibrations do a relatively poor job fitting the most metal-poor stars ([Fe/H] $< -0.5$) in the sample. Adding square terms improves the fit only negligibly ($\Delta$R$_{\mathrm{ap}}^2 < 0.02$) and does not significantly improve the results for the most metal-poor stars. This, combined with our results from Section~\ref{sec:determinemetal}, suggests that fitting metal-poor stars requires a different set of lines, rather than simply higher order terms. Improvements may also be possible by deriving a separate calibration for [Fe/H]$<-0.5$, however, our sample has only 12 stars in this range, which is insufficient to derive a reliable calibration.

We also find a calibration for determining [M/H] of the form:
\begin{eqnarray}\nonumber \label{eqn:mhkbandtest}
\mathrm{[M/H]} &=& 0.16\times \mathrm{EW}_{Na} + 0.039 \times \mathrm{EW}_{Ca}\\
&& + 2.29\times (\mathrm{H}_2\mathrm{O} \mathit{-} \mathrm{K}) - 3.04
\end{eqnarray}
which gives RMSE $= 0.13$ and R$_{\mathrm{ap}}^2 = 0.75$.


\subsection{$H$-band Metallicities}
In addition to refining the calibration of RA10, T12 derive metallicities from $H$-band spectra. The technique relies on the Ca and K lines in the $H$-band and a separate \water~band defined for the $H$-band (H$_2$O-H). As we did with $K$-band metallicities in Section \ref{sec:kband}, we use our sample to test the quality of the T12 technique. As before, we follow their prescription, including using the same continuum regions to fit the continuum to a 4th order Legendre polynomial. 

We find the T12 calibration gives RMSE $= 0.16$, and R$_{\mathrm{ap}}^2 = 0.71$. As before, we improve this calibration, and find a best fit of the form:
\begin{eqnarray}\nonumber \label{eqn:hfhfit}
\mathrm{[Fe/H]} &=&  0.55\times \mathrm{EW}_{K} + 0.32 \times \mathrm{EW}_{Ca}\\
&& + 1.1\times (\mathrm{H}_2\mathrm{O} \mathit{-} \mathrm{H}) - 2.09,
\end{eqnarray}
The new calibration gives an almost negligible improvement over T12; yielding RMSE$ = 0.14$, and R$_{\mathrm{ap}}^2 = 0.74$. 
As with the $K$-band metallicities, adding square terms improves the fit negligibly (increase in R$_{\mathrm{ap}}^2 < 0.01$), again suggesting that more lines are needed to fit the metal-poor stars.

Fitting these features to [M/H] we find a best fit of the form:
\begin{eqnarray}\nonumber\label{eqn:hmhfit}
\mathrm{[M/H]} &=&  0.41\times \mathrm{EW}_{K} + 0.24\times \mathrm{EW}_{Ca}\\
&& + 1.04 \times (\mathrm{H}_2\mathrm{O} \mathit{-} \mathrm{H}) - 1.77,
\end{eqnarray}
which gives RMSE$ = 0.12$ and R$_{\mathrm{ap}}^2 = 0.71$. 

\section{Summary and Discussion}\label{sec:discussion}
We present our sample of 112 late-K and M dwarfs in wide binary systems which we use to locate the most metal-sensitive features and recalibrate existing methods to determine late K and M dwarf metallicities. We combine published metallicities of 62 of the primary stars with 50 from our own CFHT spectra. We use moderate-resolution visible and NIR spectra of the late K and M dwarfs to identify the largest possible set of metal-dependent spectral features in late K to mid M dwarfs for each of the $JHK$ and visible wavelength bands. We utilize the metallicities of the primaries to calibrate these metal-dependent features and obtain optimal relationships to estimate metallicity in M dwarfs. Our sample covers a wide range of spectral types (from K5 to M6) and 1.5~dex in metallicity. This enables us to search for dependencies on spectral type, which was previously impossible with the relatively small samples used. We draw 5 important conclusions from our analysis: \\

1. It is possible to determine accurate (RMSE $\lesssim 0.1$~dex) metallicities for late-K to mid-M dwarfs using modest resolution spectra ($1000<R<2000$) from a variety of different wavelengths. Although features in the $K$-band perform best, metallicities can be estimated from spectra of any of the four wavelength regions.\\

2. Determining reliable metallicities at visible wavelengths requires different calibrations depending on the spectral type of the star. The results are most accurate for K5.5-M2 dwarfs, most likely because the atomic lines we use are less contaminated by molecular lines and the pseudo-continuum is easier to estimate for these dwarfs. It is not known if our calibrations are applicable for stars later than M6. This will be the subject of a future investigation on metallicities for late M and brown dwarfs.\\

3. Existing methods to determine metallicities using $H$- and $K$-band spectra (e.g. those from T12) work well for stars of near solar-metallicity, but have difficulties with most metal-poor stars in our sample ([Fe/H]$<-0.5$), even after applying our re-calibrations. Instead, determining metallicities for these stars requires the use of additional lines/features to improve the fit. This is most likely due to differences in [$\alpha$/Fe], which are not being accurately captured by the K, Ca, and Na lines (Na and K are not $\alpha$ elements) used by R10 and T12.\\

4. We are approaching the limits of what is possible with moderate resolution spectra. There is a diminishing return on adding additional lines to a given fit after 3-4 features, even if there are many more metal-sensitive features present in a given wavelength range. Thus going to higher S/N or adding more wide binaries of near solar-metallically is unlikely to improve the calibration. However, improvements could probably be made by including later spectral types (later than M5), getting more [M/H] values, a larger number of more metal-poor stars ([Fe/H]$<-1.0$), or obtaining spectra with higher resolution. \\

5. Although the $\zeta$ parameter, commonly used to place stars into metallicity classes, correctly identifies metal-poor stars used in \citet{Woolf:2009qy}, classifications based on $\zeta$ incorrectly identify 12 of the 13 K/M companions with [Fe/H]$<-0.5$ as near solar-metallicity. This suggests that the $\zeta$ parameter is sensitive to stellar characteristics other than temperature and metallicity (e.g., activity, gravity, etc.), and may incorrectly identify some metal-poor stars as having near-solar abundances.\\

Our calibrations may be useful for both existing and future catalogs of M dwarf spectra. In particular, our calibration for visible wavelength spectra can be used on existing catalogs of local M dwarfs such as \citet{Lepine:2013lr} to better probe the metallicity distribution of the local neighborhood especially since this sample is mostly early M-dwarfs, where our calibration performs best. Sloan Digital Sky Survey also has $\simeq70,000$ visible wavelength spectra of M dwarfs \citep{West:2011fj, 2011AJ....141...98B} with similar resolution to our own, which could be used in conjunction with our calibrations to map out the metallicity distribution of the sample. Work has already been done in this area to confirm the existence of an `M dwarf problem'  \citep{Woolf:2012lr}, but this depends on less metallicities derived from $\zeta$ parameter.

Although our fits have better RMSE values for [M/H] than they do for [Fe/H], this does not necessarily mean those fits are superior. In fact, the R$_{\mathrm{ap}}^2$ values for fits to [M/H] are all inferior to those calibrations done on [Fe/H]. The reason is that the distribution of values for [Fe/H] is not the same as it is for [M/H]. Some of the most metal-poor stars do not have [M/H] values, and those that do generally have higher [M/H] values due to large differences in $\alpha$ abundance (as determined for the primary star).

We confirm the claim of \citet{Johnson:2012fk}, that one can get approximate M dwarf metallicities using $J-K$ versus $V-K$ colors. However, the technique has a limited range of metallicities ($-0.4 < $[Fe/H]$<+0.2$) and is only accurate to $\simeq0.2$~dex. Thus this technique is probably best used in special applications, such as biasing a planet-search towards metal-rich M dwarfs. 

One possible explanation for the poor performance of $\zeta$ on our sample compared to that of \citet{Woolf:2009qy} is the presence of unresolved binaries. It is likely that most wide binaries form as higher order systems \citep{Kouwenhoven:2010lr}. Thus many of our wide pairs may include unresolved M+M dwarf pairs. There is evidence of radius inflation in low-mass eclipsing binary systems \citep{Lopez-Morales:2007fk,Irwin:2011lr, Kraus:2011rr} and may also be cooler than their single star counterparts \citep{Boyajian:2012lr}. Further, atmospheric models indicate that the TiO5 and CaH2/CaH3 indices, on which $\zeta$ is based, are sensitive to temperature and gravity \citep{2008AJ....136..840J,Allard:2011lr}. However, none of our most metal-poor companions show H-$\alpha$ emission, whereas radius inflation in tight binaries is usually associated with high chromospheric activity \citep{Lopez-Morales:2007fk, Kraus:2011rr, Stassun:2012uq}. Additional metallicities of M-dwarf with known multiplicity (e.g., low-mass eclipsing binaries and spectroscopic binaries) are needed to confirm if this is the source of the discrepancy.

Another complication is the possibility of having false binaries (chance alignments) in our sample. We can estimate the number of interlopers by cross referencing our sample with that of \citet{Tokovinin:2012fj}. \citet{Tokovinin:2012fj} calculate the probability that stars with commiserate proper motions are actually physically associated with each other ($P_{\mathrm{phys}}$). Although \citet{Tokovinin:2012fj} caution that their probabilities are purely based on models (and therefore only approximate), the numbers can be used to give a rough estimate of contamination from chance alignments. By summing up $P_{\mathrm{phys}}$ values for all of our binaries included in the \citet{Tokovinin:2012fj} sample we find that $>90\%$ of our binaries are physically associated with each other. However, some of the pairs with low $P_{\mathrm{phys}}$ values have parallax information for both the primary and companion that are consistent with each other. If we assume pairs with consistent parallaxes have $P_{\mathrm{phys}}=1$ and repeat our calculation, we find that $94\%$ of our binaries are physically associated with each other. Although our metal-poor stars ([Fe/H]<-0.5) tend to be more distant, and therefore only 3 of the 13 are listed in \citet{Tokovinin:2012fj}, two of them have $P_{\mathrm{phys}}>93\%$ (the other has $P_{\mathrm{phys}}=72\%$). Three more of our [Fe/H]$<-0.5$ stars have parallaxes that are consistent with the primary to 1$\sigma$, indicating that even our metal-poor stars are almost all physical pairs.

We do not claim to have identified every single metal-sensitive feature at the resolution of our spectra, however, the nature of our analysis means that it is unlikely that we missed any of the most useful ones. We perform a rough test on our recovery rate by introducing artificial metal-sensitive lines of various usefulness and repeating our analysis. Specifically, we select a sample of the most metal-sensitive features (those used in Equations (8)-(17)) and insert them elsewhere in the spectrum of the stars. When moving features from NIR to visible wavelengths we convolve the lines with a Gaussian profile to reduce the resolution of the feature (features moved into the NIR are not changed). We then repeat our analysis as described in Section~\ref{sec:MCanalysis}. We find that metal-sensitive features are sometimes not identified when they are placed on very strong telluric features, blueward of 0.4\,\um\ (where the S/N is very low), or when they overlap with other strong features (e.g., the Mg~Ib line) that make clean measurements difficult. We also note that features identified as metal-sensitive in the NIR appear less metal-sensitive (although they are still identified) when placed in visible wavelengths; this is likely due to the lower resolution and/or difficulties measuring features that are convolved with strong molecular lines in the visible. In spite of these exceptions, we still recover $>88\%$ of the lines on average, and $>93\%$ when we exclude telluric regions and low S/N regions of the spectrum. This indicates that our analysis is quite robust, and that expanding on our findings will require observations later-type stars (past M5), more metal-poor stars, or higher resolution visible spectra.

\acknowledgments
This work was supported by NSF grant AST-0908419 (Astronomy \& Astrophysics) and NNX11AC33G (Origins of Solar Systems) to EG, as well as NSF grants AST 06-07757 and AST 09-08419 to SL. We thank Andrew West for his useful suggestions this paper.

SNIFS on the UH 2.2-m telescope is part of the Nearby Supernova Factory project, a scientific collaboration among the Centre de Recherche Astronomique de Lyon, Institut de Physique NuclŽaire de Lyon, Laboratoire de Physique NuclŽaire et des Hautes Energies, Lawrence Berkeley National Laboratory, Yale University, University of Bonn, Max Planck Institute for Astrophysics, Tsinghua Center for Astrophysics, and the Centre de Physique des Particules de Marseille.

Based on observations obtained with MegaPrime/MegaCam, a joint project of CFHT and CEA/IRFU, at the Canada-France-Hawaii Telescope (CFHT) which is operated by the National Research Council (NRC) of Canada, the Institut National des Science de l'Univers of the Centre National de la Recherche Scientifique (CNRS) of France, and the University of Hawaii. This work is based in part on data products produced at Terapix available at the Canadian Astronomy Data Centre as part of the Canada-France-Hawaii Telescope Legacy Survey, a collaborative project of NRC and CNRS.

Based on observations at the Infrared Telescope Facility, which is operated by the University of Hawaii under Cooperative Agreement no. NNX-08AE38A with the National Aeronautics and Space Administration, Science Mission Directorate, Planetary Astronomy Program.

{\it Facilities:} \facility{IRTF}, \facility{CFHT},  \facility{UH:2.2m}

\bibliography{$HOME/Dropbox/fullbiblio}


\clearpage

\end{document}